\documentclass[aps,showpacs,showkeys,prc,twocolumn,groupedaddress,amsmath,amssymb,floatfix]{revtex4-1}

\usepackage{graphicx}
\usepackage{dcolumn}
\usepackage{bm}
\usepackage{rotating}

\newcommand{\dpex}{(d,p)}
\newcommand{\zrnty}{$^{90}$Zr }
\newcommand{\zrntyx}{$^{90}$Zr}
\newcommand{\zrone}{$^{91}$Zr }
\newcommand{\zronex}{$^{91}$Zr}
\newcommand{\caeight}{$^{48}$Ca }
\newcommand{\caeightx}{$^{48}$Ca}
\newcommand{\canine}{$^{49}$Ca }
\newcommand{\caninex}{$^{49}$Ca}
\newcommand{\otwen}{$^{20}$O }
\newcommand{\otwenx}{$^{20}$O}
\newcommand{\oone}{$^{21}$O }
\newcommand{\oonex}{$^{21}$O}
\newcommand{\bea}{\begin{eqnarray}}
\newcommand{\eea}{\end{eqnarray}}

\begin{document} 

\title{Revisiting surface-integral formulations for one-nucleon transfers to bound and resonance states}

\author{J.E. Escher$^{(a)}$} 
\email{escher1@llnl.gov}
\author{I.J.~Thompson$^{(a)}$}
\author{G.~Arbanas$^{(b)}$}
\author{Ch.~Elster$^{(c)}$}
\author{V.~Eremenko$^{(c,e)}$}
\author{L.~Hlophe$^{(c)}$}
\author{F.M.~Nunes$^{(d)}$}
\affiliation{(a) Lawrence Livermore National Laboratory L-414, Livermore, CA 94551, USA}
\affiliation{(b) Reactor and Nuclear Systems Division, Oak Ridge National Laboratory, Oak Ridge, TN 37831, USA}
\affiliation{(c) Institute of Nuclear and Particle Physics,  and Department of Physics and Astronomy,  Ohio University, Athens, OH 45701}
\affiliation{(d) National Superconducting Cyclotron Laboratory and Department of Physics and Astronomy, Michigan State University, East Lansing, MI 48824, USA}
\affiliation{(e) D.V. Skobeltsyn Institute of Nuclear Physics, M.V. Lomonosov Moscow State University, Moscow, 119991, Russia}
\collaboration{The TORUS Collaboration }
\noaffiliation

\date{\today}

\begin{abstract}
One-nucleon transfer reactions, in particular (d,p) reactions, have played a central role in nuclear structure studies for many decades.  Present theoretical descriptions of the underlying reaction mechanisms are insufficient for addressing the challenges and opportunities that are opening up with new radioactive beam facilities.  We investigate a theoretical approach that was proposed recently to address shortcomings in the description of transfers to resonance states.  The method builds on ideas from the very successful R-matrix theory; in particular it uses a similar separation of the parameter space into interior and exterior regions, and introduces a parameterization that can be related to physical observables, which, in principle, makes it possible to extract meaningful spectroscopic information from experiments. 
We carry out calculations, for a selection of isotopes and energies, to test the usefulness of the new approach. 
\end{abstract}



\pacs{24.10.-i, 25.45.Hi, 24.30.-v}
\keywords{Transfer reactions, resonance reactions}
\maketitle


\section{Introduction}
\label{sec_intro}

Deuteron-induced reactions, in particular (d,p) one-neutron transfer reactions, have been used for decades to investigate the structure of nuclei. These reactions, carried out in inverse kinematics, are expected to play a central role in the study of weakly-bound systems at modern radioactive beam facilities~\cite{Jones:10}. While the theoretical framework and its computational implementation for describing (d,p) reactions have seen much progress over the decades, open questions remain and need to be addressed.  Resonances, for example, occur in all nuclei, and are of special interest in the low-energy spectra of weakly-bound nuclei.
Their properties, in particular their energies and widths, are of great interest to nuclear astrophysics, as many reaction that occur in astrophysical environments proceed through resonances. Furthermore, resonances provide both challenges and stringent tests for nuclear structure models, as predicting their properties involves a proper treatment of the nuclear many-body system including the continuum~\cite{Volya:06a}.

Current theoretical descriptions of transfer reactions that populate resonance states suffer from two major shortcomings. On the practical side, one has to deal with numerical convergence issues, as the matrix element that describes the transition to the final state includes contributions from very large distances outside the nucleus.
Conceptually, it is also not clear what spectroscopic information can be extracted from a transfer to a resonance. A clear connection between experimental transfer observables and typical resonance parameters remains to be established.  

Recently, a new formalism that utilizes concepts known from the successful and popular R-matrix theory was proposed for the description of deuteron-induced reactions~\cite{Mukhamedzhanov:11a}. 
R-matrix theory is a standard tool for extracting resonance properties such as energies and widths from nucleon capture and scattering experiments~\cite{Lane:58,Descouvemont:10}, and the new approach establishes a similar link between resonance properties and transfer reactions.
The formalism covers transfers to bound and resonance states, and is general enough to include deuteron breakup. 

A central tenet of the formalism proposed in Ref.~\cite{Mukhamedzhanov:11a} is the recasting of the reaction amplitude, a volume integral, in terms of a surface integral plus (presumably small) remnant terms that contain contributions from the interior and exterior of the final nucleus. 
Interior and exterior are defined with respect to the distance between the transferred nucleon and the target nucleus.
The surface-integral formalism, as we will refer to it here, holds the potential to overcome present difficulties in describing transfers to resonance states and to become a practical method for extracting structure information from transfer experiments, since: 1) it reduces the dependence of the cross section calculations on the model used for the nuclear interior; 2) it avoids the convergence problems that affect traditional calculations of transfers to resonances; and 3) it establishes a useful link between resonance properties and transfer observables, since the cross section obtained from the surface integral can be parameterized in terms of quantities that are familiar from traditional R-matrix approaches.

It is the purpose of this paper to test the formalism proposed in Ref.~\cite{Mukhamedzhanov:11a}. In particular, we study the role of interior and exterior contributions to the cross sections, for both post and prior formulations of the transfer amplitudes. This allows us to draw some conclusions about the sensitivity of the post and prior expressions to model assumptions made about the nuclear interior.  We then investigate the relative contributions from the surface term and the residual interior and exterior terms to the transfer cross section and the dependence of these contributions on the chosen radius $a$ at which the surface integral is evaluated.  
This is done in order to assess to which extent the surface term captures the essentials of the reaction and reproduces the full cross section.
The calculations are carried out in the framework of the DWBA formalism. The implications for generalizing the study to the Continuum-Discretized Coupled-Channels (CDCC) formalism~\cite{Austern:87}, which incorporates effects of deuteron breakup in the reaction, are briefly considered.

In the next section, Section~\ref{sec_formalism}, we summarize the formalism developed in Ref.~\cite{Mukhamedzhanov:11a}, in particular the introduction of `exterior' and `interior' contributions to the transfer matrix element and the emergence of a surface term. In Section~\ref{sec_study}, we investigate the contributions of the interior, surface, and exterior terms to the transfer cross sections. Our findings are summarized in Section~\ref{sec_summary}.

\section{Review of the Formalism}
\label{sec_formalism}

In the R-matrix approach for a binary reaction~\cite{Lane:58,Descouvemont:10}, the configuration space is separated into two regions: an exterior region, in which the reaction partners are well-separated and interact only via the Coulomb interaction (for charged particles), and an interior region, where a confined compound system exists that is governed by nuclear and Coulomb forces. Formally, the nuclear wave function in the interior is expanded in some suitable set of basis functions, while in the exterior, it takes the form of a scattering wave function, and matching conditions are imposed at the surface. In typical applications, the parameters (resonance energies and width parameters) are adjusted to reproduce measured cross sections. This approach makes it possible to parameterize the collision matrix, and thus the calculated cross section, in terms of a few formal R-matrix parameters, which can then be related to observed quantities.
R-matrix theory~\cite{Lane:58} is typically employed in the description of binary reactions, such as elastic or inelastic scattering, or capture reactions, but the underlying ideas can be applied more generally.  

For reactions of the type $A$(d,p)$F$, where $A$ denotes the target and $F$ the residual nucleus, Mukhamedzhanov~\cite{Mukhamedzhanov:11a} introduced an analogous separation of the model space into interior and exterior regions.  The separation into the different regions is based on the distance $r_{nA}$ of the deposited neutron from the center of the target nucleus. 
The standard distorted-wave Born Approximation (DWBA) transition matrix element can be written in post or prior form~\cite{Satchler:Book}:
\begin{eqnarray}
M^{(post)} &=& \left< \varphi_F \chi_{pF}^{(-)} | \Delta V_{pF} | \varphi_d \varphi_A \chi_{dA}^{(+)} \right> \nonumber \\
&=& \left< I_A^F \chi_{pF}^{(-)} | \Delta V_{pF} | \varphi_d  \chi_{dA}^{(+)} \right> \label{eq:MPost} \\
M^{(prior)} &=& \left< \varphi_F \chi_{pF}^{(-)} | \Delta V_{dA} | \varphi_d \varphi_A \chi_{dA}^{(+)} \right>  \nonumber \\
&=& \left< I_A^F \chi_{pF}^{(-)} | \Delta V_{dA} | \varphi_d  \chi_{dA}^{(+)} \right> \label{eq:MPrior}
\end{eqnarray}
where $\varphi_A$ and $\varphi_F$ denote the wave functions of the initial ($A$) and final ($F = A+n$) nuclei, respectively, and $I_A^F (r_{nA})$ $=\left< \varphi_A | \varphi_F \right>$ is the associated overlap function, which depends on the coordinate of interest, $r_{nA}$.
In applications, this is typically approximated by a single-particle wave function obtained from a potential-model calculation~\cite{Escher:01PRC, Escher:02PRC}.
The distorted waves in the entrance and exit channels are given by $\chi_{dA}$ and $\chi_{pF}$, respectively, and $\varphi_d$ is the deuteron wave function. 

Post and prior forms require different transition operators
\begin{eqnarray}
\Delta V_{pF} &=& V_{pn} + V_{pA} - U_{pF}, \; \mbox{(post)} \label{eq_transOp_post} \\
\Delta V_{dA} &=& V_{nA} + V_{pA} - U_{dA},  \; \mbox{(prior)} \label{eq_transOp_prior}
\end{eqnarray}
respectively. In the post form, $\Delta V_{pF}$ contains the interaction between the proton and the target nucleus ($V_{pA}$), the proton-neutron interaction ($V_{pn}$), and the optical potential for the exit channel ($U_{pF}$), while in the prior form,  $\Delta V_{dA}$ contains the interaction between the neutron and the target ($V_{nA}$), the proton and the target ($V_{pA}$), and the optical potential in the entrance channel ($U_{dA}$).  The interactions include both nuclear and Coulomb terms.  
In the first-order distorted-wave approach, post and prior formulations give identical matrix elements, provided the interactions and wave functions are chosen consistently~\cite{Satchler:Book}.  This implies, e.g., that the deuteron wave function $\varphi_d$ has to be an eigenfunction of the potential $V_{pn}$ (prior form), and that the overlap function $I_A^F$ has to be an eigenfunction of $V_{nA}$ (post form). Similarly, $U_{dA}$ and $U_{pF}$ have to be consistent with the distorted waves $\chi_{dA}$ and $\chi_{pF}$, respectively.

While post and prior forms are equivalent, historically there has been a preference for using the post formalism for describing (d,p) reactions.  One advantage of using the transition operator in Eq.~\ref{eq_transOp_post} over using the one in Eq.~\ref{eq_transOp_prior} is that it lends itself to the approximation $V_{pA} - U_{pF} \approx 0$ and $V_{pn}$ is short-ranged, which makes it possible to treat the interaction between proton and neutron inside the deuteron in a zero-range approximation~\cite{Satchler:Book}.  In addition, the Johnson-Soper model~\cite{Johnson:70}, an early description of deuteron stripping and elastic scattering that includes the effects of deuteron breakup, is formulated in the post form and makes use of the zero-range approximation. Similarly, the Johnson-Tandy~\cite{Johnson:74} extension of the model to finite range uses the post form. 
One disadvantage of the post form, however, is that for stripping reactions that populate resonance final states, convergence becomes difficult. Unlike in the prior case, where contributions from the long tails of resonance wave functions are suppressed by the short-range nature of the interactions in Eq.~\ref{eq_transOp_prior}, post calculations typically have contributions from the resonance wave function at very large distances $r_{nA}$, due to the $V_{pn}$ term.

In both the post and prior formulations, the transition matrix element can be separated into two parts,
\begin{eqnarray}
M^{(DWBA)} = M_{int}(0,a) + M_{ext}(a,\infty), 
\label{eq_intext}
\end{eqnarray}
where $a$ refers to the specific value chosen for the coordinate $r_{nA}$, for which to carry out the separation.
The interior term involves an integration from $r_{nA}=0$ to $r_{nA}=a$, while the exterior term involves an integration from $r_{nA}=a$ to very large radii, $r_{nA} \rightarrow \infty$.
In Section~\ref{sec_study}, we will illustrate that for typical (d,p) transfer reactions and separation radii roughly equal to the size of the
target nucleus, the main contributions to the post form matrix element come from the exterior region, while the prior form matrix element is dominated by interior contributions from further inside the nucleus. This is significant, as in Ref.~\cite{Mukhamedzhanov:11a}, it was further demonstrated that Green's Theorem can be employed to convert the post-form exterior matrix element into a surface integral plus a prior-form exterior matrix element:
\begin{eqnarray}
\label{eq_extPoToSurfPlusPri}
M^{(post)}_{ext}(a,\infty) &=&  M_{surf}(a) + M^{(prior)}_{ext}(a,\infty).
\end{eqnarray}

Using this in the post version of Eq.~\ref{eq_intext}, the full DWBA matrix element becomes:
\begin{eqnarray}
\label{eq_intsurext}
\label{eq_MSurf}
\lefteqn{M^{(DWBA)} =} \\
&& M^{(post)}_{int}(0,a) + M_{surf}(a) + M^{(prior)}_{ext}(a,\infty). \nonumber
\end{eqnarray}
The same result is obtained if one starts with a consideration of the prior form and expresses the interior part in terms of a surface integral and remaining contributions from the interior, 
\begin{eqnarray}
M_{int}^{(prior)}(0,a) &=& M_{surf} (a) + M_{int}^{(post)}(0,a) .
\label{eq_intPoToSurfPlusPri}
\end{eqnarray}

The interior post term in Eq.~\ref{eq_MSurf} involves an integration over the overlap function $I_A^F (r_{nA})$ for small $r_{nA}$, {\it i.e.\ } it depends on a model for the nuclear interior.  
The exterior prior and surface terms, on the other hand, are related to the asymptotic properties of the overlap function, as long as $a$ is outside the boundary of the neutron binding potential.
  
For transfer reactions that populate resonance reactions, this surface formulation is particularly attractive.
In Ref.~\cite{Mukhamedzhanov:11a}, it was shown that the surface term $M_{surf}(a)$, which is evaluated at a specific radius $a$, can be parameterized by quantities that are familiar from traditional R-matrix approaches, namely a channel radius (here the separation radius $a$), logarithmic boundary conditions (here logarithmic derivatives of known Hankel functions), and reduced-width amplitudes (here related to the asymptotic normalization of the overlap function).
Thus, a useful link between resonance properties and transfer observables could be established.  

A dominant surface matrix element would reduce the dependence of the cross section calculations on the model used for the interior portion of the overlap function. 

In the next section we investigate the relative contributions of interior, surface, and exterior terms, for various target nuclei and at various reaction energies.  We will focus on the DWBA implementation.

\section{Interior, surface, and exterior contributions to the transfer cross section}
\label{sec_study}

Transfer cross sections calculated in the DWBA framework are proportional to the square of the full matrix element $M^{(DWBA)}$ given in Eqs.~\ref{eq_intext} and \ref{eq_intsurext}.
To better assess the relevance of the various contributions for realistic situations, we begin with a full cross section calculation, with model parameters selected to reasonably reproduce measured (d,p) cross sections.  
We then repeat each cross section calculation with all contributions other than the term of interest set to zero and compare the resulting cross section to the full calculation.

This comparison is made in one of two ways: 1) We compare the peak cross section obtained in the restricted calculation to that obtained in the full calculation.  This is done when we are interested in the dependence of the contribution on the separation radius $a$.  In that case we plot the ratio 
\begin{eqnarray}
R_{X(a)} &=& \frac{\hat{\sigma}^{(DWBA)}_{X(a)}}{\hat{\sigma}^{(DWBA)}_{exact}}  \; ,
\label{eq_ratio}
\end{eqnarray}
where $\hat{\sigma}^{(DWBA)}_{exact}$ denotes the peak value of the full cross section, and $\hat{\sigma}^{(DWBA)}_{X(a)}$ is the peak value of the cross section obtained by including only the term $X(a)$, for $X(a) = $ $M_{int}(0,a)$, $M_{surf}(a)$, or $M_{ext}(a,\infty)$.
We consider the first (smallest-angle) peak, as this peak plays a prominent role in comparisons of calculations with experimental data.
2) For specific cases of interest, we also compare the angular behavior of the cross section $\sigma^{(DWBA)}_{X(a)}$ to $\sigma^{(DWBA)}_{exact}$.
Focusing in particular on $\sigma_{M_{surf}(a)}^{(DWBA)}$  allows us to assess to which extent the surface term is able to describe the reaction considered.

In our study, we included target nuclei from different mass regions and considered different beam energies. 
Here, we present representative calculations for the target nuclei $^{20}$O, $^{48}$Ca, and $^{90}$Zr.  Additional cases ($^{12}$C, $^{40}$Ca, $^{208}$Pb) give similar results.  
For the $^{90}$Zr case, we considered transfer reactions with $E_d$ = 11 MeV~\cite{Michelman:69}. We calculated cross sections for transfers to the $5/2^+$ ground state, the $1/2^+$ first excited state, and to a narrow $f_{7/2}$ resonance state that lies about 1 MeV above the neutron separation threshold of 7.195 MeV.
For  $^{48}$Ca, we studied three different beam energies, $E_d$ = 13, 19.3, and 56 MeV~\cite{Metz:75,Uozumi:94}, and considered reactions involving both the $3/2^-$ ground and first excited $1/2^-$ state in $^{49}$Ca.
The $^{20}$O(d,p) reaction was recently studied in inverse kinematics, at equivalent deuteron energy $E_d$ = 21 MeV. Measurements were carried out for transfers to two bound states ($5/2^+$ ground state and $1/2^+$ first excited state) and two resonance states (a $3/2^+$ resonance at  $E_{ex}$ = 4.77 MeV and a resonance at $E_{ex}$ = 6.17 MeV, which has either $3/2^+$ or $7/2^-$ character.) We carried out calculations for the two bound states and the two resonances, considering both spin-parity assignments for the second resonance.

Finite-range DWBA calculations were carried out with a modified version of the reaction code {\sc Fresco}~\cite{Thompson:88} that allowed setting specific contributions to the transition matrix element to zero.
The Reid Soft-Core potential~\cite{Reid:68} was used to describe the deuteron.
We used the deuteron-nucleus optical-model potential by Daehnick~\cite{Daehnick:80} and employed the Koning-Delaroche potential~\cite{Koning:03} for the proton-nucleus interaction. 
The manner in which we calculated the surface term, as a difference between prior-interior and post-interior contributions (Eq.~\ref{eq_intPoToSurfPlusPri}), requires good post-prior agreement.
We therefore matched the approximation used by the {\sc Fresco} code, specifically the omission of the spin-orbit term in $U_{dA}$ and $U_{pF}$ in the transition operators (Eqs.~\ref{eq_transOp_post} and \ref{eq_transOp_prior}), by eliminating this term also from the optical-model potential in the entrance and exit channels.
In order to still carry out comparisons to experimental results, small adjustments to the potential parameters were made to reproduce the elastic scattering cross section, where available, and to improve the fit of the full calculations with measured (d,p) cross sections.  We verified that the overall conclusions of our study are not affected by these changes in potential parameters.
For final bound states, the overlap function was approximated by a single-particle function in a Woods-Saxon well, with a depth adjusted to reproduce the proper binding energy.
For resonance states, we adopted a bin description for the overlap function.
Bin functions are momentum-averaged continuum wave functions $\Phi(r) =$ $\sqrt{\frac{2}{\pi N}}$ $\int_{k_1}^{k_2} w(k) \phi_k(r) dk$ (with $N$ = $\int_{k_1}^{k_2} |w(k)|^2 dk$) that are square integrable, orthogonal to bound states and to each other (provided the momentum ranges do not overlap)~\cite{Thompson:88}. Here $\phi_k(r)$ is a positive-energy scattering state of the potential. If the radius of integration over $r$ is taken to be sufficiently large, the bin functions are normalized and the transfer calculations converge. For the narrow resonances considered here, we used a weight function of the form $w(k) =$ exp($-i\delta_k$)sin($\delta_k$), where $\delta_k$ is the scattering phase shift for $\phi_k(r)$.

We first consider the separation of the full matrix element into an interior and an exterior contribution, as given in Eq.~\ref{eq_intext}. Subsequently, we study the separation into three terms, given in Eq.~\ref{eq_intsurext} and discuss the role of the surface term relative to the other contributions.

\subsection{Interior and exterior contributions: Post vs. prior formalism}
\label{subsec_results_intExt}

Transfer experiments are often carried out under conditions that presumably make the reaction peripheral, {\it i.e.\ } they probe the overlap function in the region of the nuclear surface.
The primary contributions to the transfer matrix elements should therefore come from $r_{nA}$ values that are roughly equal to the size of the target nucleus.  
Our calculations, however, will illustrate that the importance of a particular nuclear region depends on the form (post or prior) chosen. 
The potential well used for calculating the one-neutron overlap functions of various states in \zrone with the ground state of \zrnty had a radius of 5.53 fm and a diffuseness of 0.66 fm;  the rms radius of the overlap associated with the $5/2^+$ \zrone ground state was 4.94 fm.

\begin{figure}[h]
\includegraphics[width=0.7\columnwidth,angle=270,trim=80 20 40 75,clip=true]{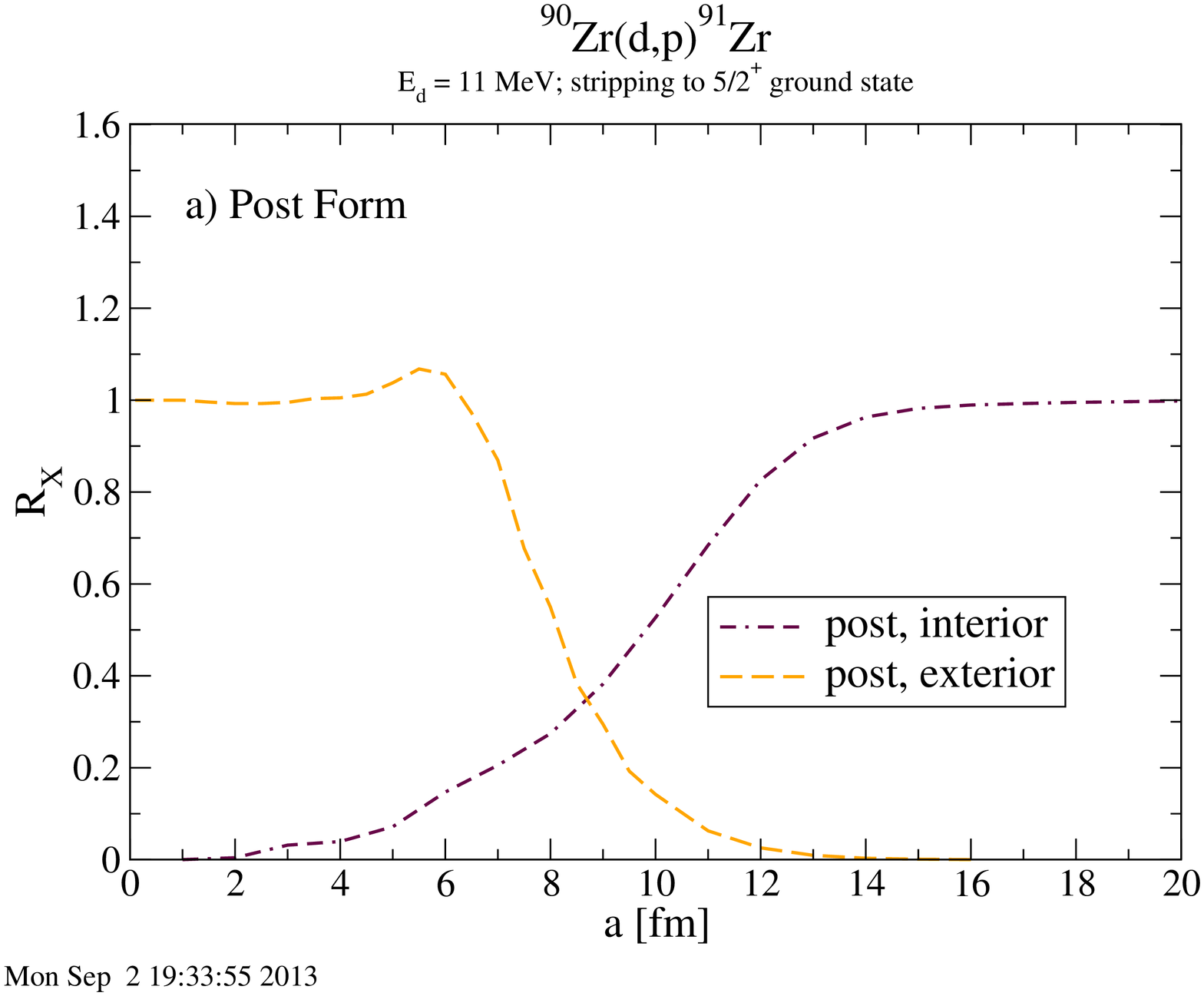}
\includegraphics[width=0.7\columnwidth,angle=270,trim=80 20 40 75,clip=true]{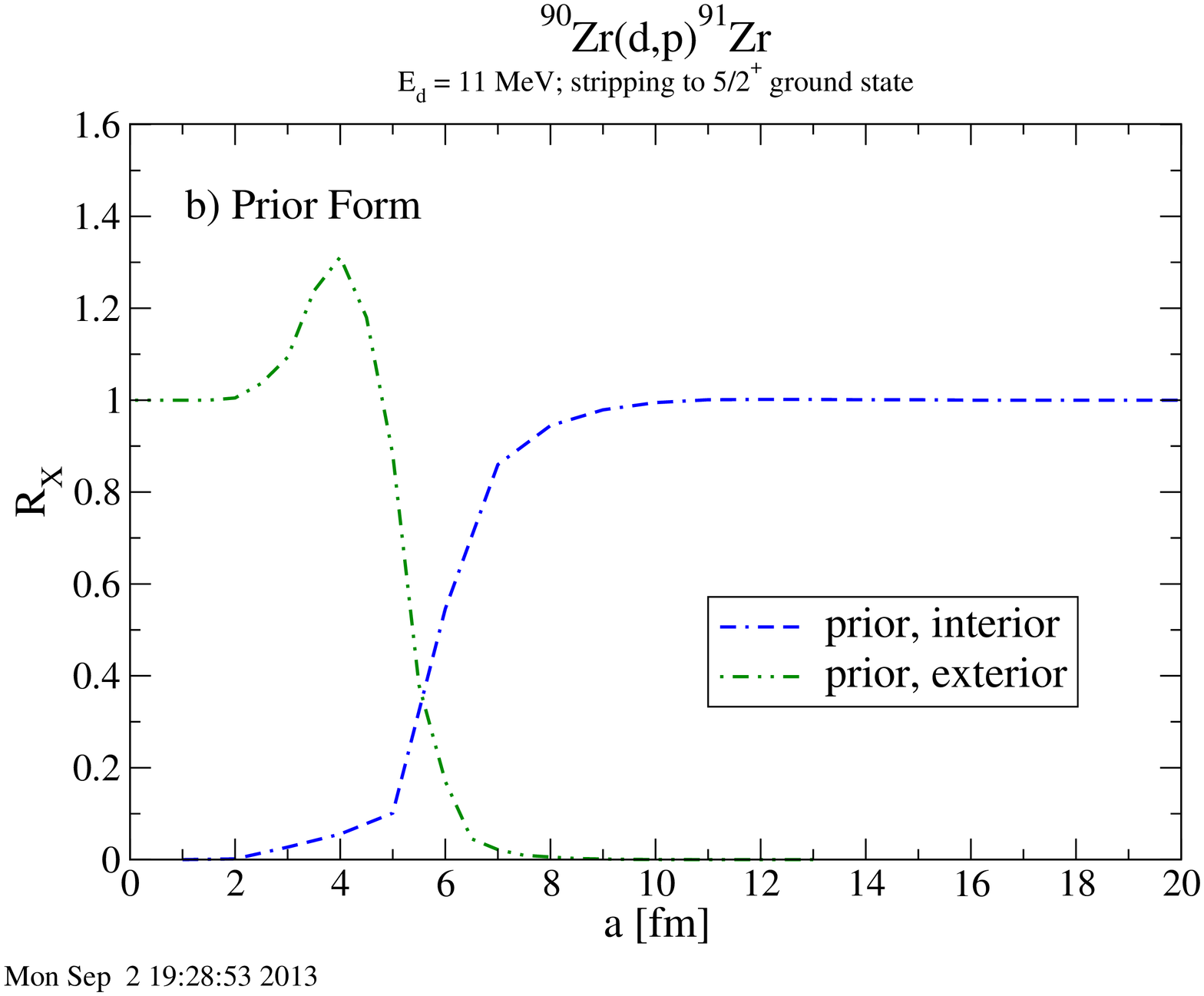}
\caption{\label{fig_intContribs}
(Color online)
Examination of the role of interior and exterior contributions for $^{90}$Zr(d,p)$^{91}$Zr stripping to the ground state. 
Shown is $R_{X(a)}$, the ratio of the peak cross section obtained when only one term (either $M_{int}$ or $M_{ext}$) is included in the transition matrix element in Eq.~\ref{eq_intext} to the peak cross sections obtained in the full calculation. The quantity is given as a function of the separation radius $a$, for both the post (a) and prior (b) formalisms.
}
\end{figure}

Figure~\ref{fig_intContribs} shows the contributions to the transfer cross section arising from the interior and exterior terms in Eq.~\ref{eq_intext}.
Results for the post and prior forms are given in the top and bottom panels, respectively.
The ratio $R_{X(a)}$ of the calculated peak cross section to the peak cross section obtained in the full calculation, is given as a function of the separation radius $a$. 
As expected, for small separation radii, the exterior terms contain all cross section contributions ($R_{M_{ext}(a)}$ $\rightarrow 1$ for $a \rightarrow 0$) and for very large radii, the interior terms carry the bulk of the cross section ($R_{M_{int}(a)}$ $\rightarrow 1$ for $a \rightarrow \infty$).  In both post and prior cases, contributions to the transition matrix element are very small for radii below 2 fm as well as for radii above 15 fm, as indicated by curves that are largely flat in those regions.
The interesting region, however, is where the cross-over from interior-dominated to exterior-dominated occurs.  We observe that this cross-over lies at much smaller radii when the prior formulation is used than in the post formulation.

In the post formulation, integration over $r_{nA}$ from 0 out to 7-8 fm yields less than 30\% of the peak cross section, while in the prior formulation such integration captures over 80\% (see post and prior {\em interior} contributions).
Similarly, when considering exterior contributions in the post form, it suffices to integrate from 4-5 fm on out, while in the prior form contributions from about 2 fm out have to be included.
At the same time, the region beyond $r_{nA} \approx $ 7-8 fm contributes little in the prior form.

This illustrates that, although the full post and prior DWBA amplitudes are equal, their behavior is quite
different in the subspace over the variable $r_{nA}$. The prior formulation is clearly more sensitive to
the nuclear interior (and thus to model assumptions about the interior structure), while contributions
from more peripheral parts of the n + target system dominate the post form. 
(The radial shape of the overlap function can be seen in the upper portion of Figure~\ref{fig_Zr_bsPostPriorSurface_3States}b.)
The reason for this behavior can be found in the difference in the structure of the transition operators, Eqs.~\ref{eq_transOp_post} and \ref{eq_transOp_prior}. 
For sufficiently large $r_{nA}$, the three terms in $\Delta V_{dA}$ will cease to contribute to the prior transition matrix element, while $V_{pn}$ in $\Delta V_{pF}$ can still produce non-negligible contributions to the post form ($V_{pA}$ and $U_{pF}$ approximately cancel each other).
It is the presence of $V_{pn}$ in the post-form matrix element that makes convergence of the radial integrals difficult when the overlap function $I_A^F$ describes a resonance~\cite{Huby:65,Vincent:70}.

Calculations for (d,p) reactions on other nuclei (C, O, Ca, Pb) show similar trends for bound states. For resonance states we find reduced contributions from the nuclear interior, but additional complications arise, as will be discussed in Section~\ref{subsec_results_surf}.

\subsection{Role of the surface term}
\label{subsec_results_surf}

In order to assess  whether it is possible to approximate the (d,p) transfer cross section by a calculation based solely on the surface term, it is necessary to investigate the contributions of the three terms in Eq.~\ref{eq_intsurext} to the full cross section. 

To calculate the terms $M^{(post)}_{int}(0,a)$ and $M^{(prior)}_{ext}(a,\infty)$, respectively, we set the overlap function to zero for $r_{nA}$ $> a$ in the first case and for $r_{nA}$ $< a$ in the second case. The surface term is calculated in first-order DWBA as $M_{surf} (a)$ = $M_{int}^{(prior)}(0,a)$ - $M_{int}^{(post)}(0,a)$. We discuss our findings next.

\subsubsection{\zrntyx\dpex\zronex: Transfers to bound and resonance states}
\label{sec_results_surf_Zr}

\begin{figure*}[htb]
\begin{minipage}[b]{0.48\textwidth} 
\begin{center}
\includegraphics[width=0.7\columnwidth,angle=270,trim=80 25 30 70,clip=true]{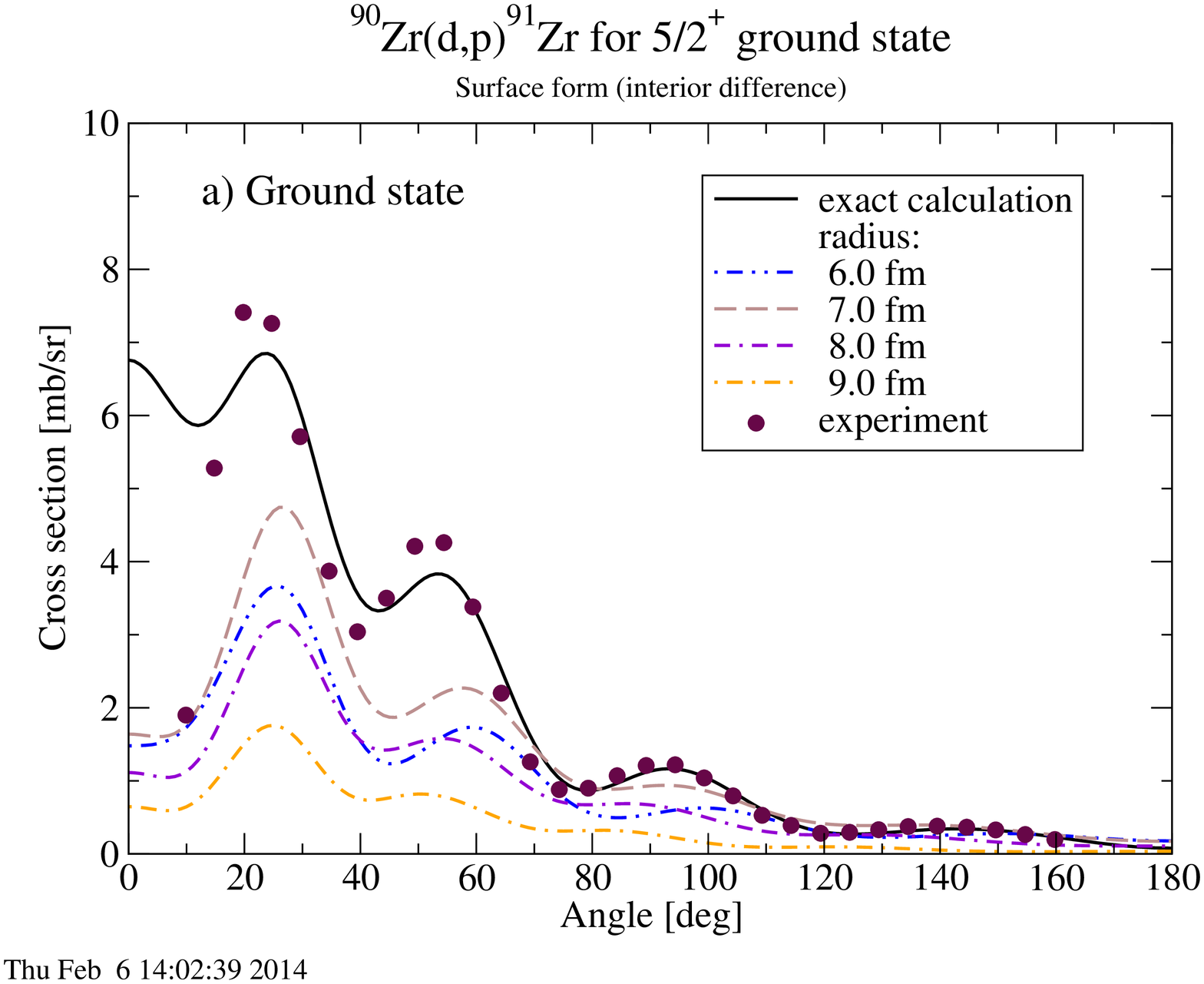}
\includegraphics[width=0.7\columnwidth,angle=270,trim=80 25 30 70,clip=true]{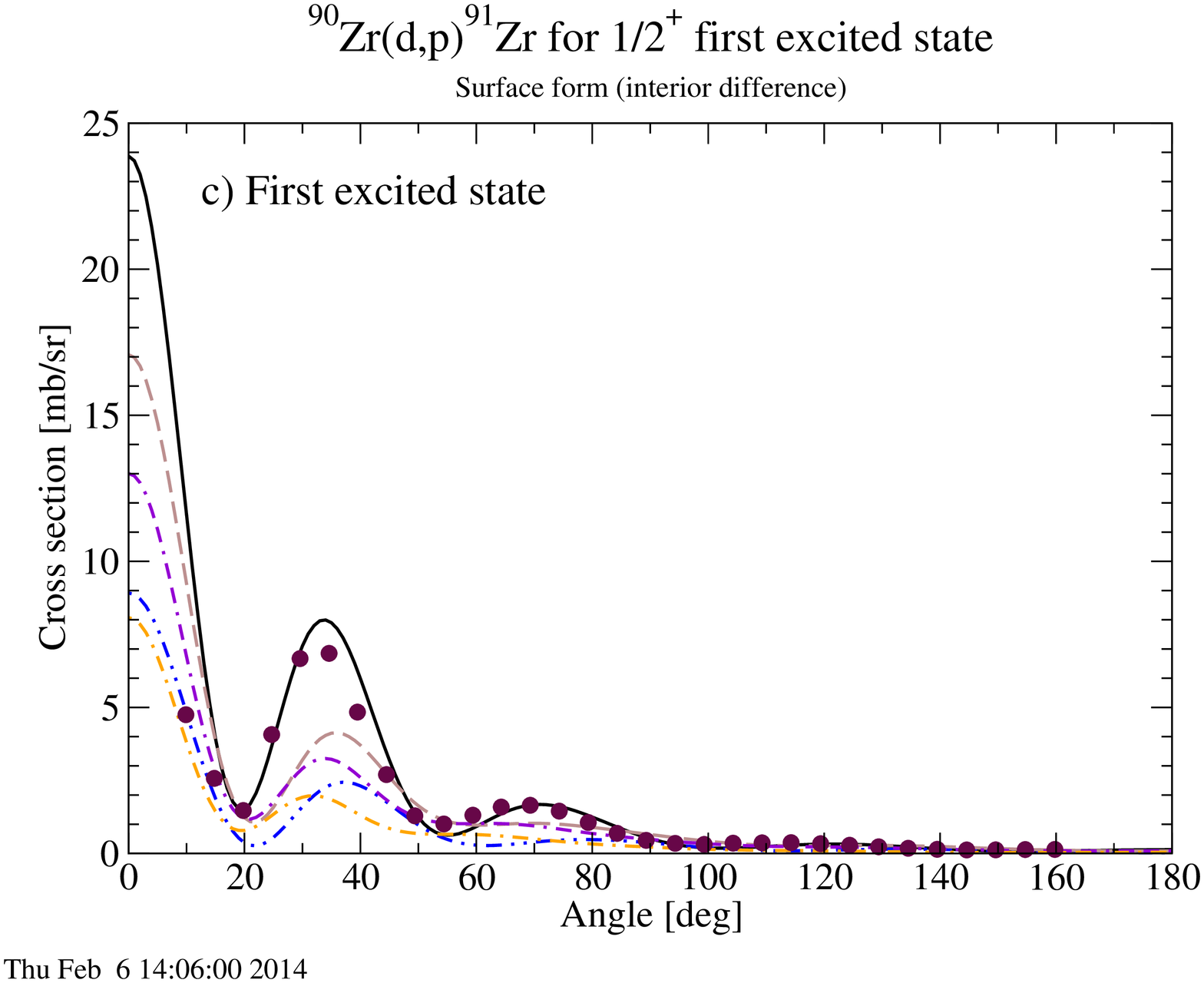} 
\includegraphics[width=0.7\columnwidth,angle=270,trim=80 25 30 70,clip=true]{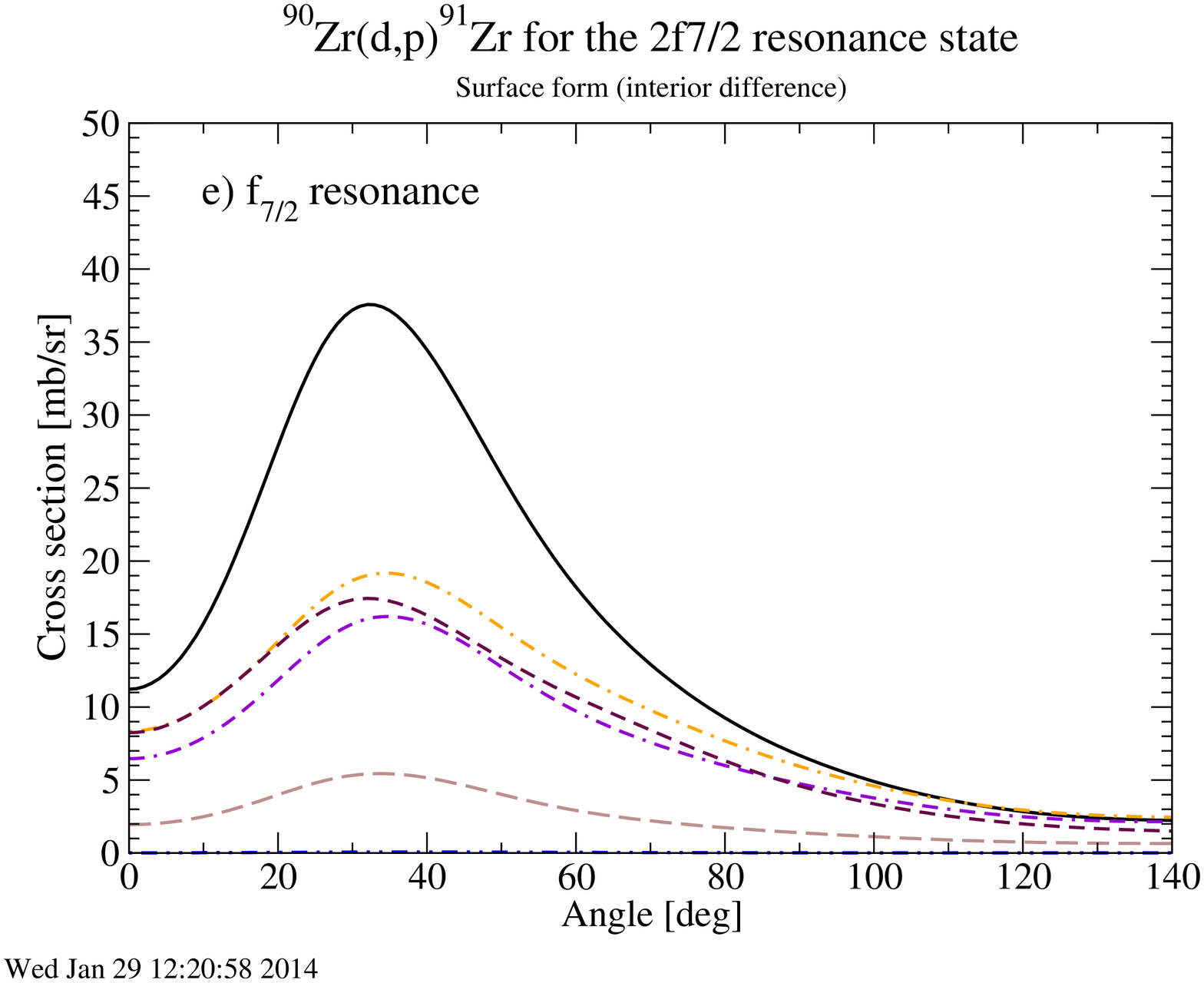} 
\end{center}
\end{minipage}
\hspace{0.1cm}
\begin{minipage}[b]{0.48\textwidth} 
\begin{center}
\includegraphics[width=0.7\columnwidth,angle=270,trim=80 18 30 40,clip=true]{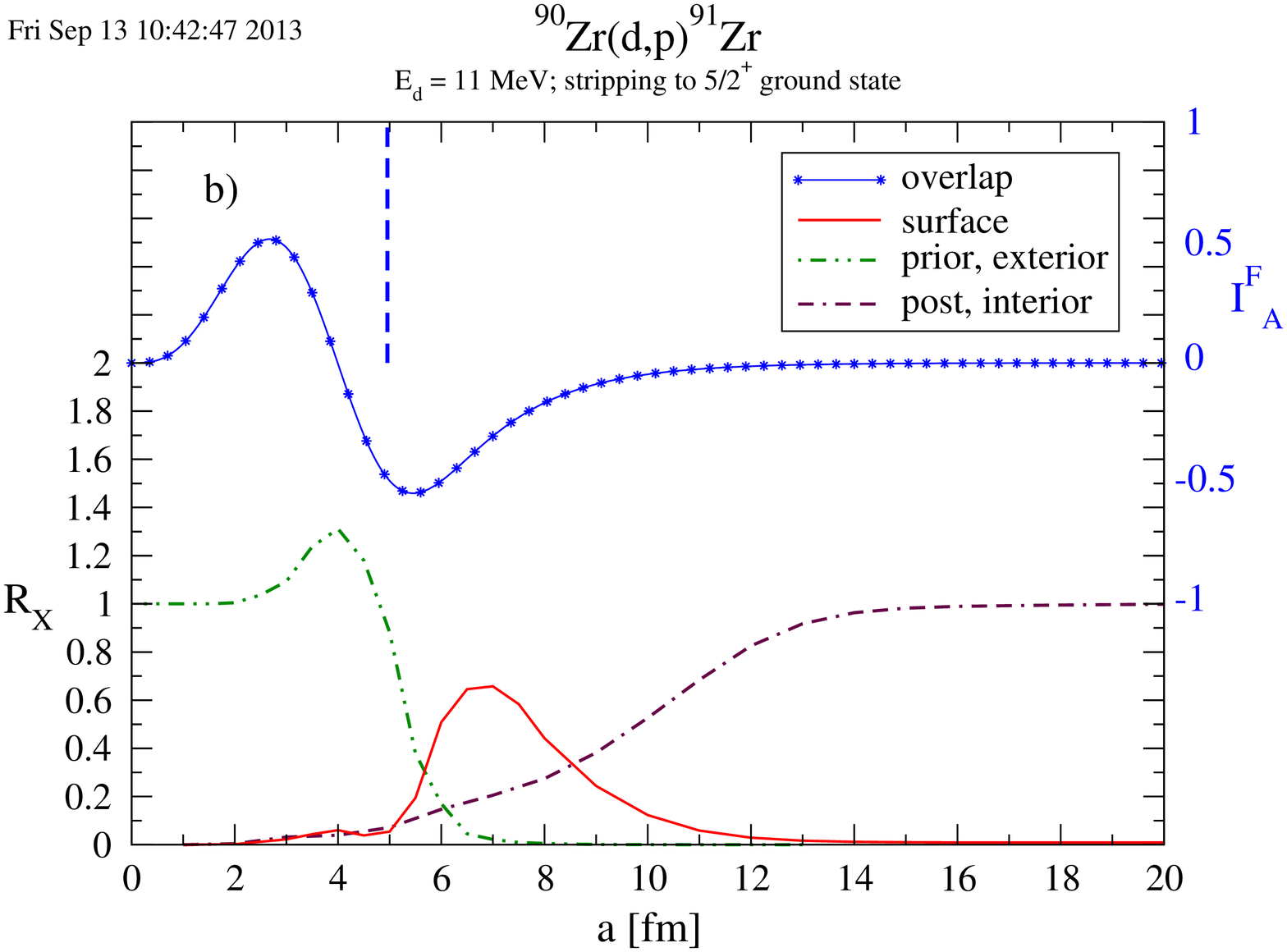}
\includegraphics[width=0.7\columnwidth,angle=270,trim=80 18 30 40,clip=true]{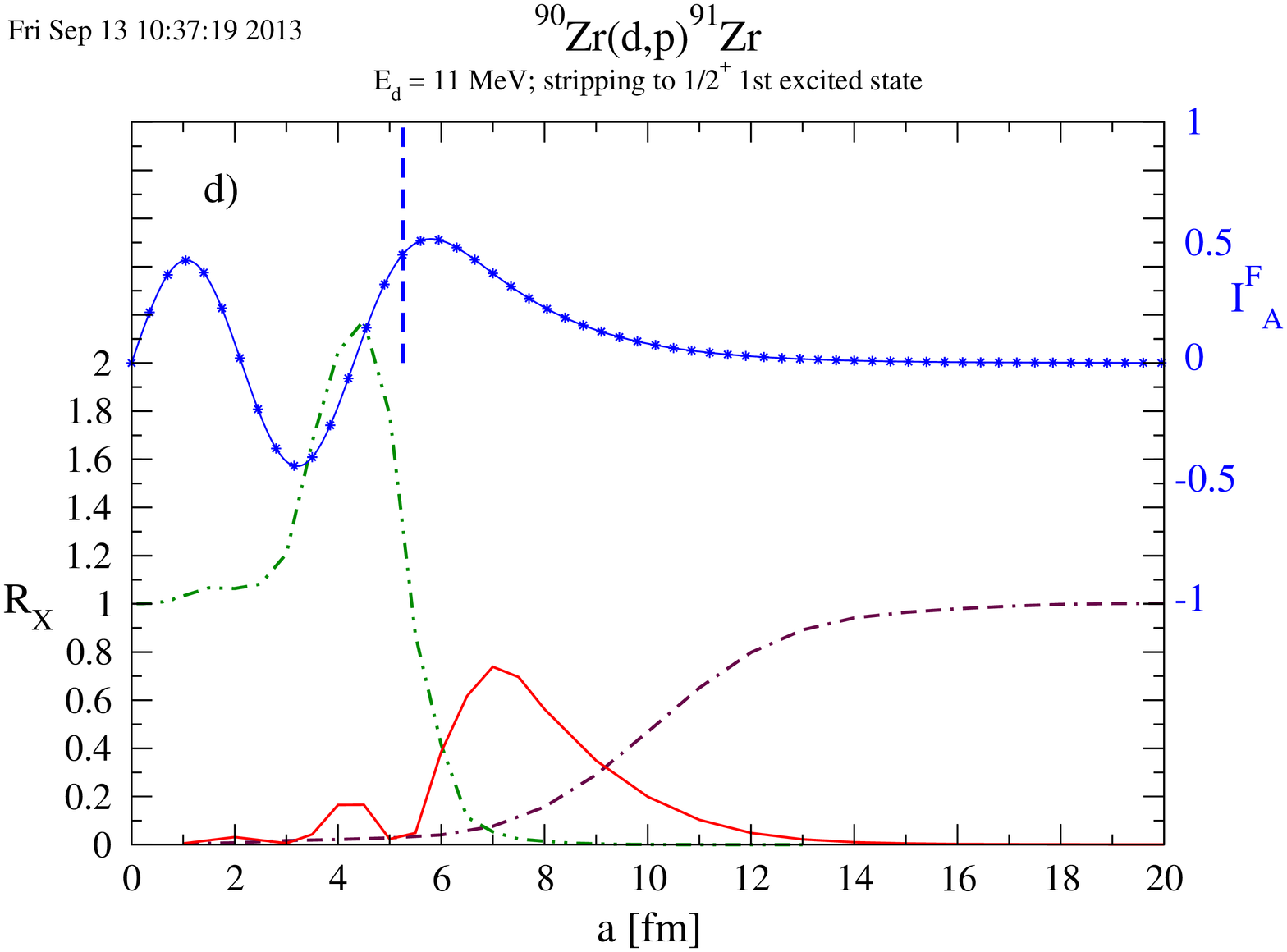}
\includegraphics[width=0.7\columnwidth,angle=270,trim=80 18 30 40,clip=true]{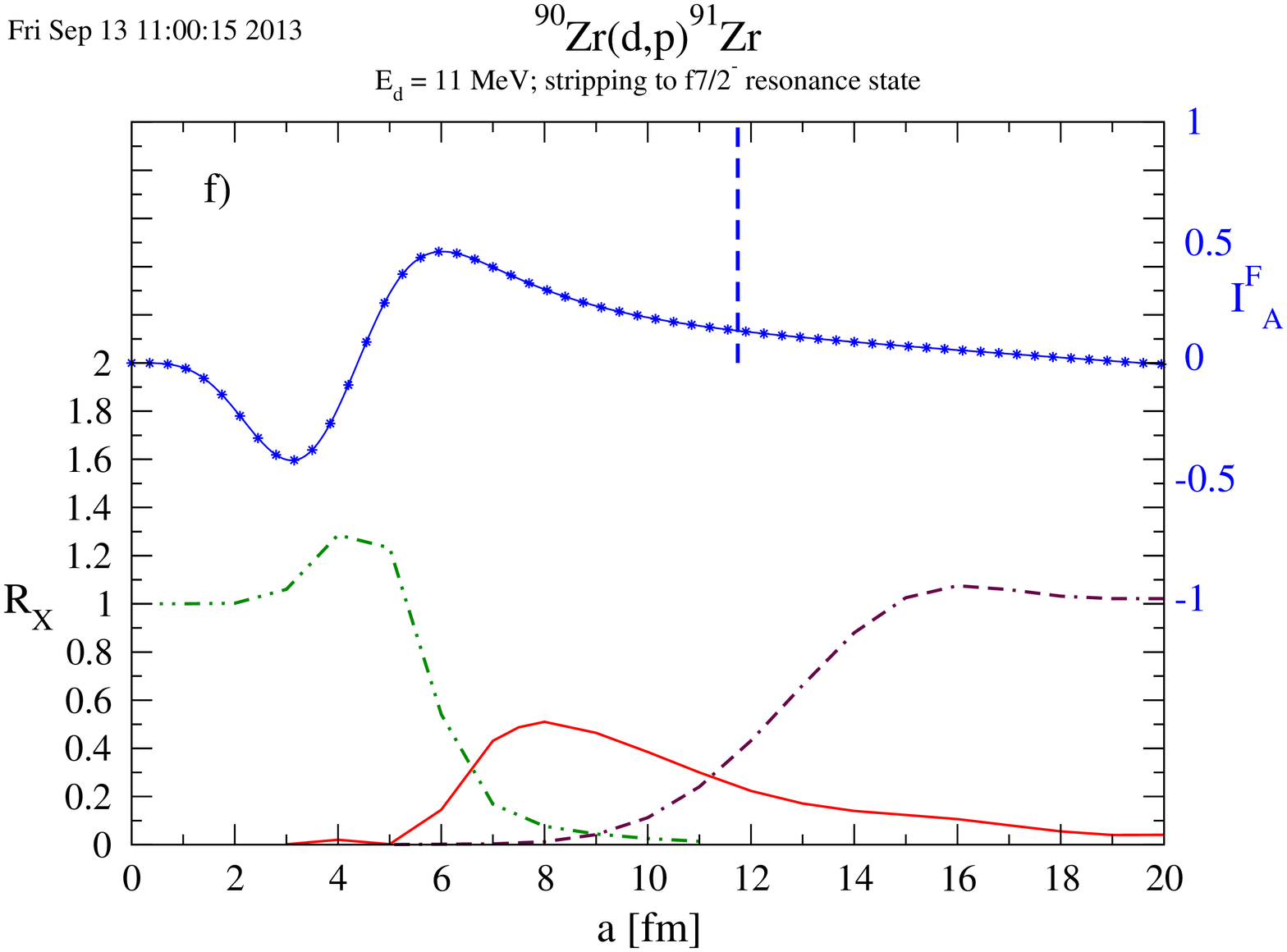} 
\end{center}
\end{minipage}
\caption{
(Color online)
Examination of interior, surface, and exterior contributions to the transfer reaction \zrntyx\dpex\zronex. Compared are several calculations for stripping to the $5/2^+$ ground state (top panels) the $1/2^+$ first excited state (middle panels), and the $f_{7/2}$ resonance (bottom panels).
Left column (panels a, c, e): Angular cross sections calculated from the surface term only (broken curves) are compared to the full stripping cross section (solid black line) and experimental data from Ref.~\cite{Michelman:69}. The broken curves illustrate the dependence of the cross section on the choice of the surface radius $a$.
Right column (panels b, d, f): Fractions $R_{X(a)}$ for the interior (post) term, the surface term, and the exterior (prior) term, are given as a function of the surface radius, in the lower portion of each panel. The ratios take values between zero and $\approx2.2$ (labels are on the left axis). The associated one-neutron overlap functions $I$ are shown in the upper portion of each panel (solid curves with stars). They are given in units of fm$^{-1}$, labels are on the right axis.  The rms radii of the overlaps are indicated by dashed vertical lines. The potential $V_{nA}$, which binds the neutron to the \zrnty  nucleus, has a radius of 5.53 fm and a diffuseness of 0.66 fm.
}
\label{fig_Zr_bsPostPriorSurface_3States}
\end{figure*}

We carried out a set of calculations for \zrntyx\dpex\zrone that included stripping to the (bound) ground and first excited states, as well as to a narrow $f_{7/2}$ resonance at about 1 MeV above the neutron separation threshold.
The panels in the right column of Figure~\ref{fig_Zr_bsPostPriorSurface_3States} show the ratios $R_{X(a)}$ of the peak cross sections obtained in the restricted calculations to the peak cross section from the full calculation, as a function of the separation radius $a$.
Also shown is the radial behavior of the one-neutron overlap functions $I$ (solid curve with stars) for the three final states considered; their rms radii are indicated by dashed vertical lines.

We find that for bound final states, the surface term is dominant at around 6-8 fm, at radii a little larger than the rms values of the overlap functions. For the resonance case, panel f), we observe that the surface term is dominant at slightly larger radii, around 7-10 fm, even though the overlap function has an rms value that is still larger, namely 11.7 fm. 
In addition, the surface term in the resonance case is non-zero over a wider range of radii than it is in the bound-state case. 

We also observe that the surface term does not fully account for the cross section. At all separation radii for which $M_{surf}(a)$ dominates, there are also remnant contributions from one or both of the other terms, $M_{int}(0,a)$ and $M_{ext}(a,\infty)$. As a result, the surface term alone does not completely reproduce the angular behavior of the cross section. This can be seen in the left column of the figure, where the cross sections calculated by using the surface terms only are compared to the full cross sections (solid curves) for selected separation radii $a$
\footnote{
For transfer to the \zrone ground state, there is a clear discrepancy in shape between the full calculations and the cross sections calculated with the surface term, with the latter not exhibiting the upturn at small angles.  In fact, in the region below 10 degrees, the measured cross section does not exhibit this upturn. The full calculation is not able to reproduce the experimental data correctly without introducing a (somewhat artifical) lower cutoff in the integration.  This issue has already been discussed in the literature, see, e.g., Ref.~\cite{Hodgeson:book}, p. 456.
}.
The calculated cross sections show some dependence on the choice of $a$, and the absence of the other terms is clearly significant.

\subsubsection{$^{48}$Ca(d,p)$^{49}$Ca: Energy dependence of the surface term}
\label{sec_results_surf_Ca}

\begin{figure*}[htb]
\begin{minipage}[b]{0.48\textwidth} 
\begin{center}
\includegraphics[width=0.7\columnwidth,angle=270,trim=80 25 30 70,clip=true]{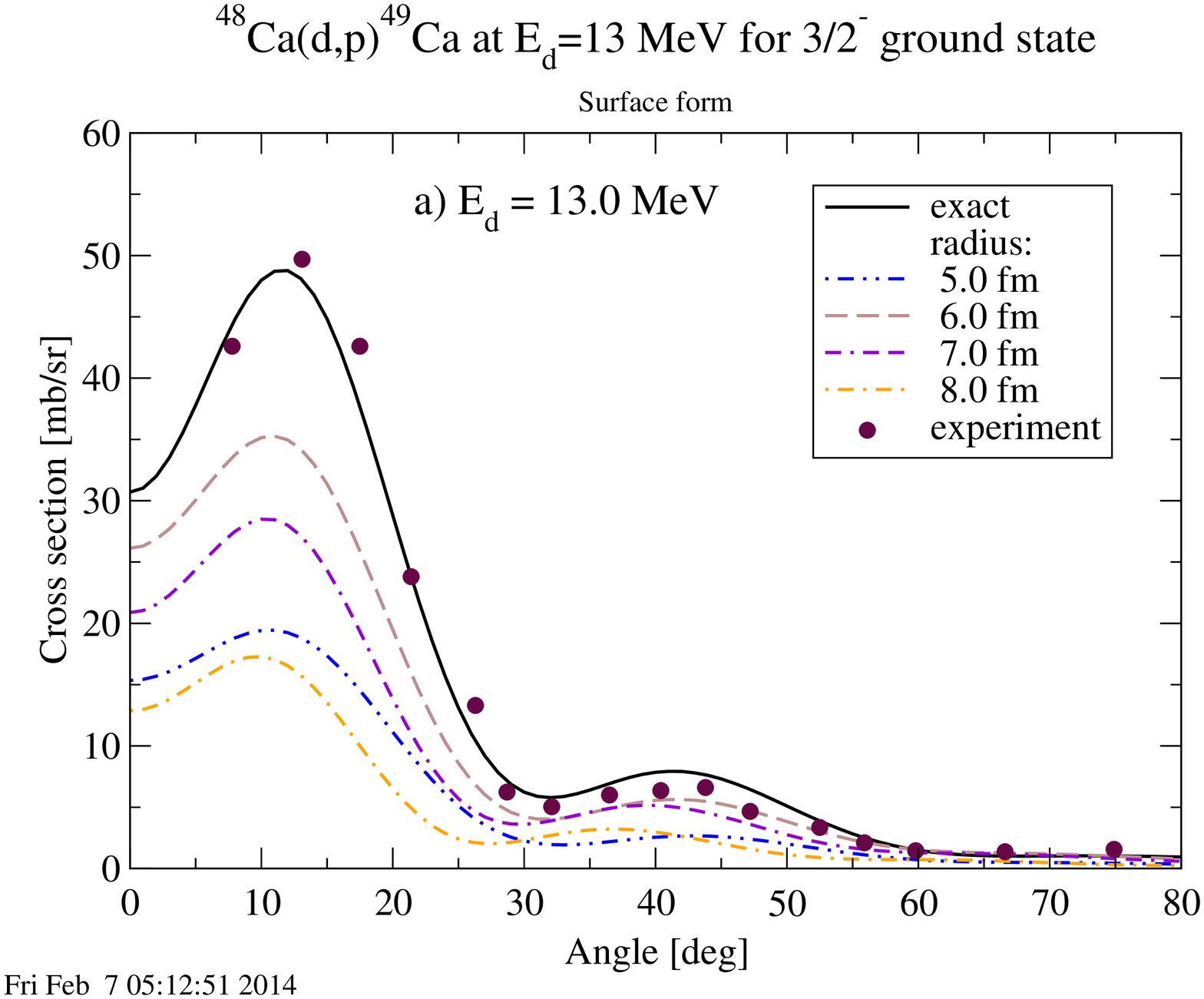}
\includegraphics[width=0.7\columnwidth,angle=270,trim=80 25 30 70,clip=true]{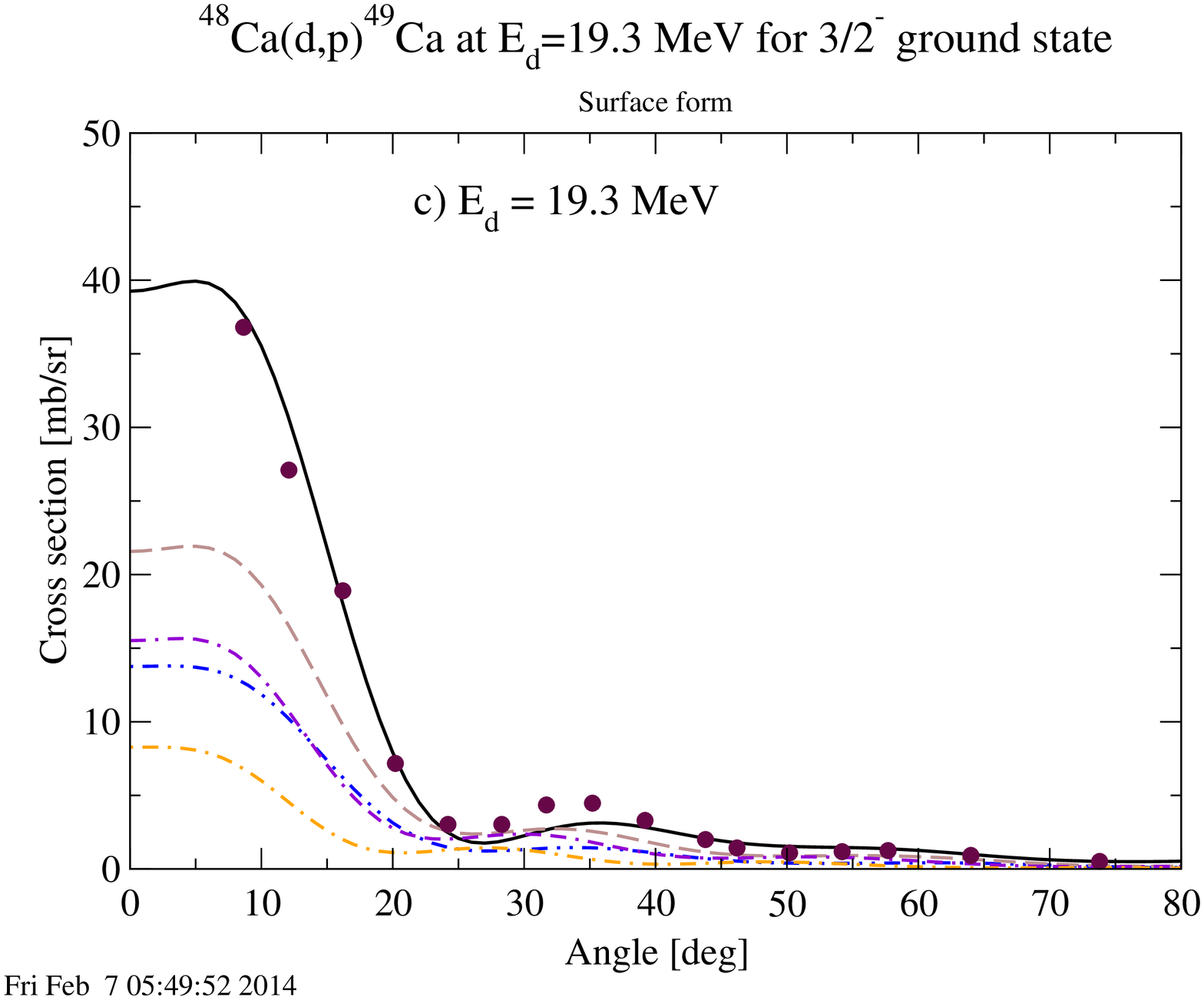} 
\includegraphics[width=0.7\columnwidth,angle=270,trim=80 25 30 70,clip=true]{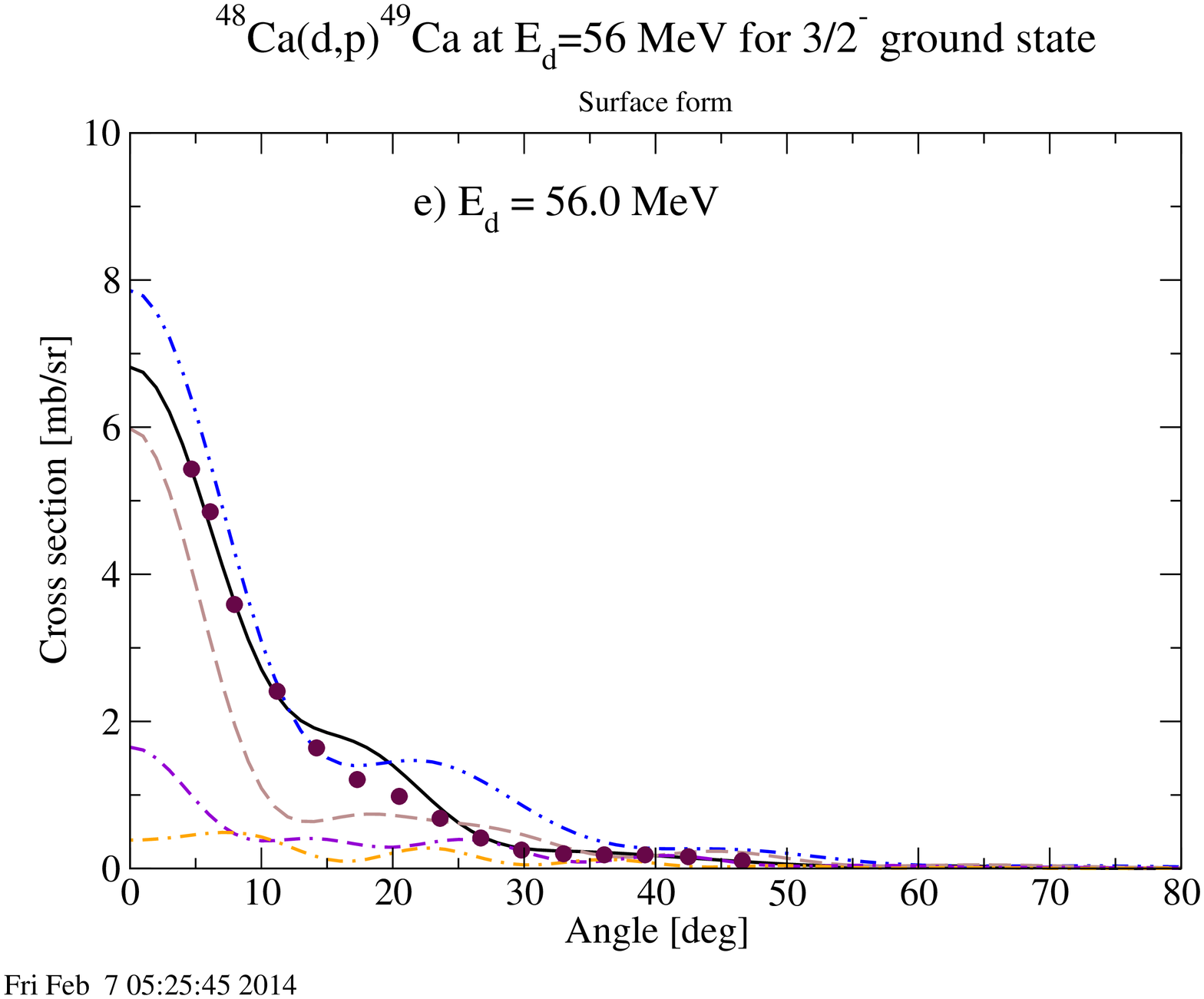}
\end{center}
\end{minipage}
\hspace{0.1cm}
\begin{minipage}[b]{0.48\textwidth} 
\begin{center}
\includegraphics[width=0.7\columnwidth,angle=270,trim=80 25 30 70,clip=true]{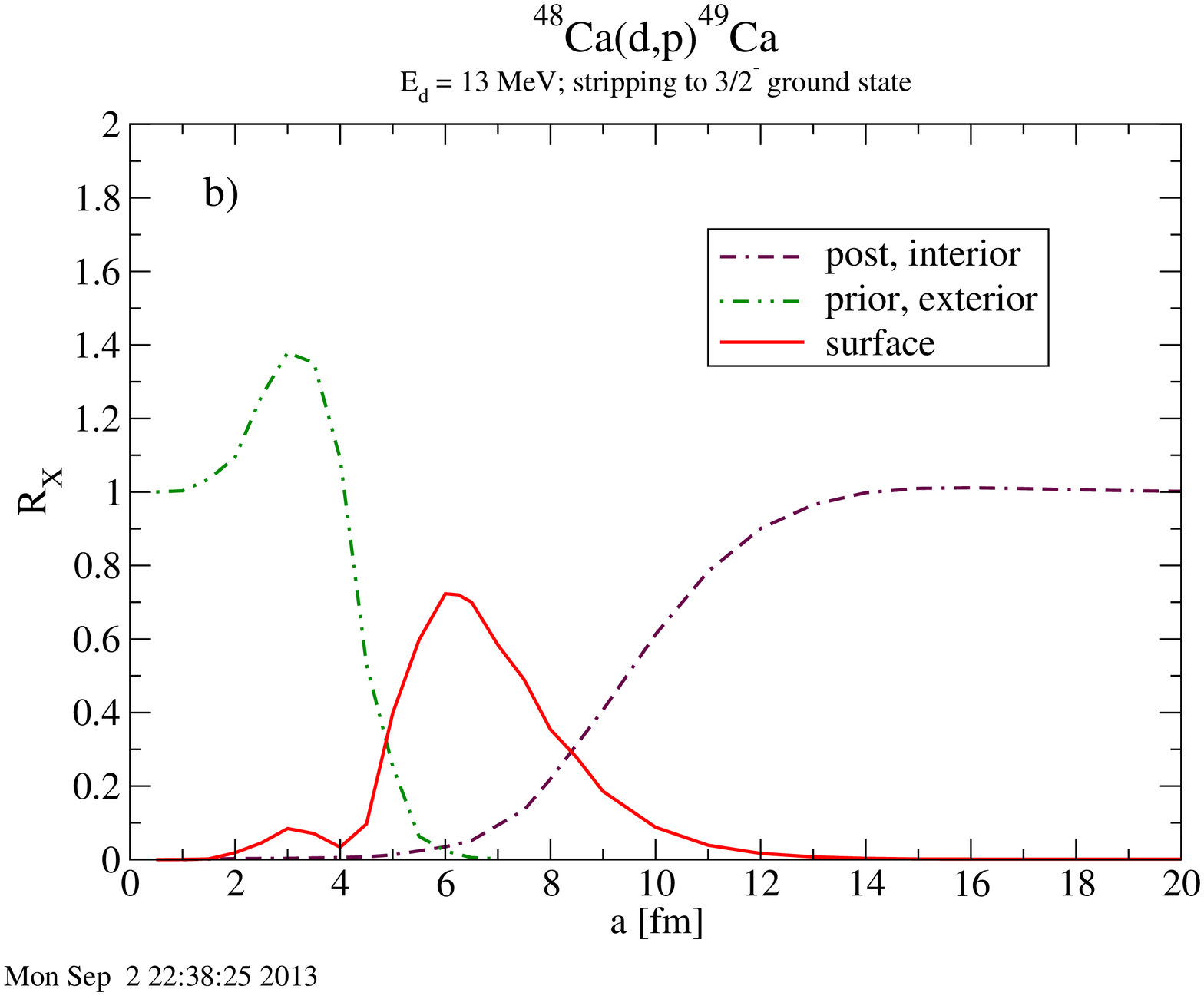}
\includegraphics[width=0.7\columnwidth,angle=270,trim=80 25 30 70,clip=true]{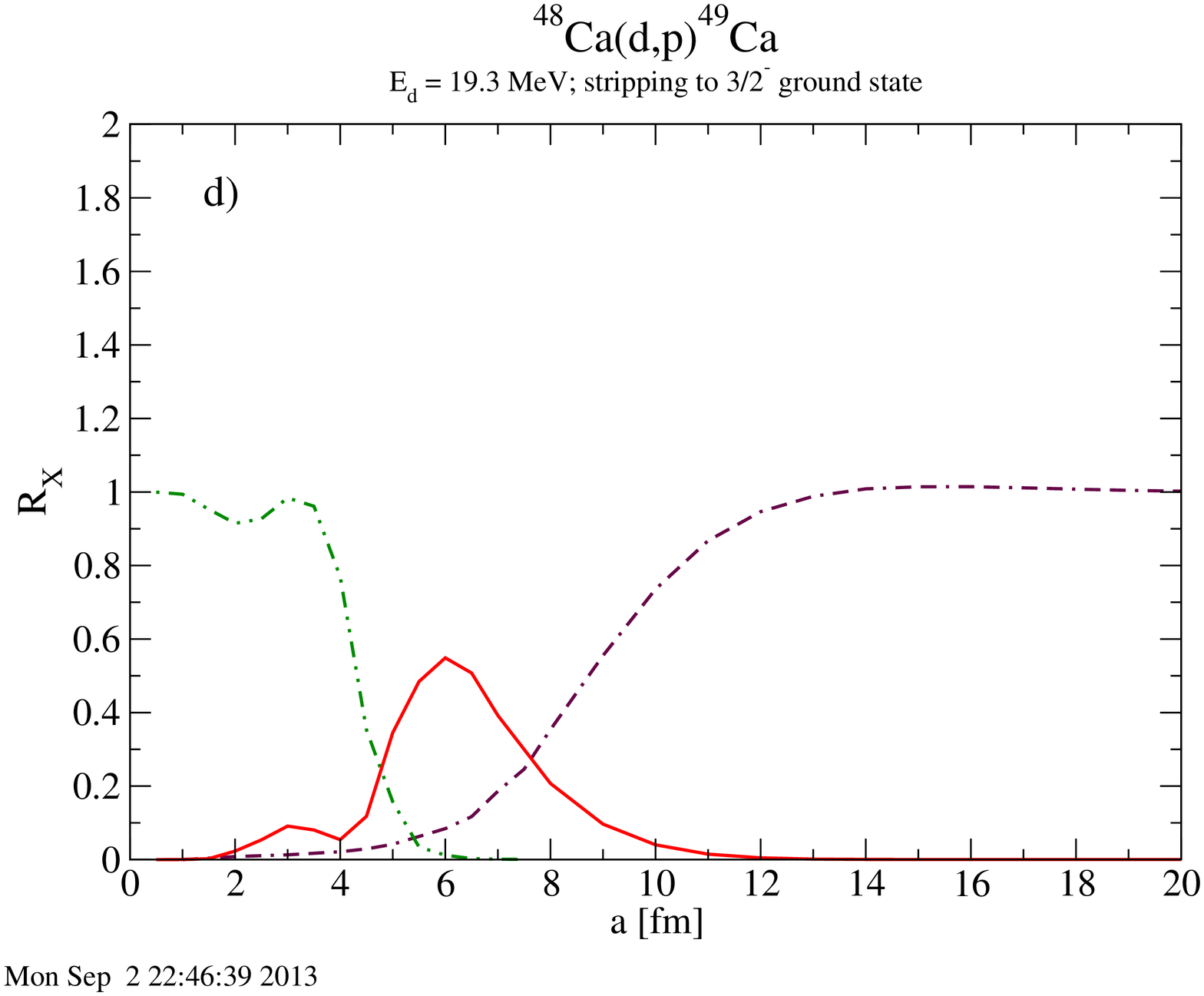}
\includegraphics[width=0.7\columnwidth,angle=270,trim=80 25 30 70,clip=true]{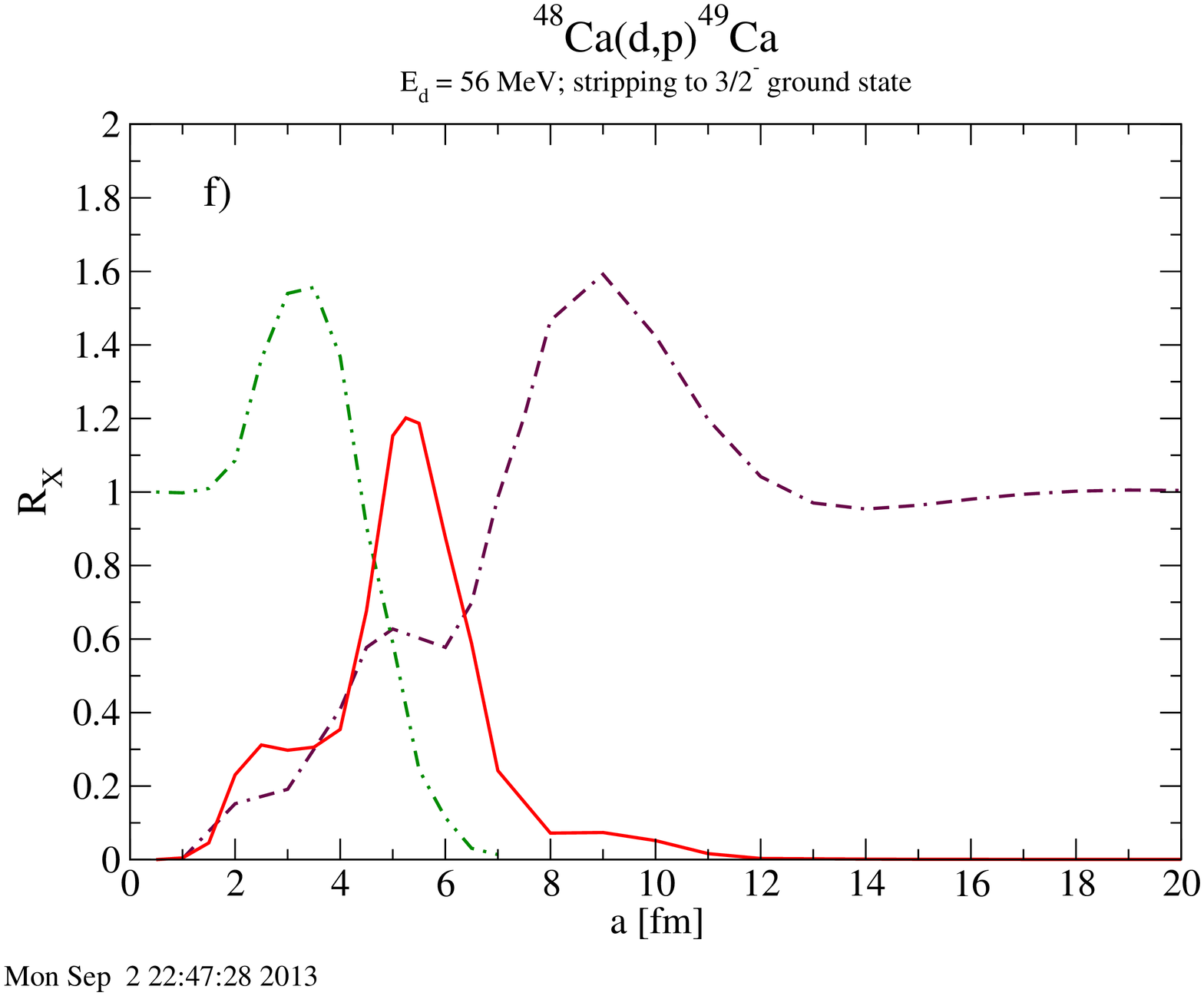}
\end{center}
\end{minipage}
\caption{
(Color online)
Examination of interior, surface, and exterior contributions for  \caeightx\dpex\caninex. Compared are several calculations for stripping to the $3/2^-$ ground state, at $E_d$ = 13, 19.3, 56 MeV  (top to bottom). 
Left column (panels a, c, e): Contributions from the surface term (broken curves) are compared to the full stripping cross section (solid black line), for various choices of the surface radius. Data are from Metz {\it et al.\ }~\cite{Metz:75} (13, 19.3 MeV) and Uozumi {\it et al.\ }~\cite{Uozumi:94} (56 MeV).
Right column (panels b, d, f): contributions resulting from the interior (post) term, the surface term, and the exterior (prior) term, are given as a function of the surface radius.  The rms radius of the associated one-neutron overlap function is 4.62 fm. 
The potential $V_{nA}$, which binds the neutron to the  \caeight nucleus, has a radius of 4.54 fm and a diffuseness of 0.65 fm. 
Results for transfers to the $1/2^-$ first excited state in \canine are very similar, but are not shown here.
}
\label{fig_Ca_bsPostPriorSurface_gs}
\end{figure*}

It is known that with increasing beam energy a reaction becomes less peripheral and is increasingly affected by the structure of the nuclear interior.  This effect is visible in Figure~\ref{fig_Ca_bsPostPriorSurface_gs}, where results are presented for \caeightx\dpex\caninex, for populating the ground state of \caninex, at three different beam energies, $E_{b}$ = 13.0, 19.3, and 56 MeV.
With increasing energy the region where the transition from exterior-dominated to interior-dominated occurs, and where the surface term dominates, shifts to smaller radii.

The first two cases resemble the Zr example discussed, with a dominant surface term around $a$ = 6.5 fm and 7.0 fm, respectively, that approximately reproduces the shape of the full cross section but underpredicts the magnitude.  
For the highest beam energy considered, $E_d$ = 56 MeV, however, the surface term peaks at smaller radii, around $a$ = 5.25 fm. In this region, there are sizable contributions from the post-interior and the prior-exterior terms. The angular behavior of the cross section for this case also confirms that at this energy, the surface term does not provide a good approximation to the full cross section. There is a strong dependence of the shape of the cross section on the separation radius chosen.  Calculations carried out for transfers to the first excited state in \canine give very similar results (not shown).  In both cases, the surface term does not provide a good representation of the reaction for the higher energies, but shows clear improvements as one moves to lower energies, $E_d$ = 19.3 and 13.0 MeV: The shape of the calculated cross section becomes less dependent on the choice of of the surface radius $a$ and better approximates the exact cross section.

\subsubsection{$^{20}$O(d,p)$^{21}$O: Focus on resonances }
\label{sec_results_surf_O}

Achieving convergence for transfer calculations involving resonances has long been known to be difficult and we find this again in the calculations presented here. 
Numerical convergence is an issue, ambiguities in how to extract structure information (resonance energies and widths) is another.
Various methods have been suggested to achieve numerical convergence~\cite{Huby:65,Fortune:69}, but their implementation in modern reaction codes is incomplete.  
Here we investigate to which extent the surface term is able to reproduce the full cross sections for several resonances.
A surface term that captures the essentials of the reaction would reduce the dependence of the calculated cross section on slowly-converging radial integrals. Furthermore, it would provide strong motivation for the implementation of the formalism within the framework of the Continuum-Discretized Coupled-Channels (CDCC) approach, as the prior term is identically zero in that case~\cite{Mukhamedzhanov:11a}.

We focus on the  \otwenx(d,p)\oone reaction, which was recently investigated in inverse kinematics with a radioactive beam~\cite{Fernandez:11}. This is one of very few cases where angular cross sections for both bound and resonance states have been measured in the same experiment. Using a 10.5/u MeV \otwen beam, it was possible to obtain angular cross sections for one-nucleon transfers populating the bound ground and first excited states, as well as resonances at 4.77 MeV and 6.17 MeV (lying 0.96 MeV and 2.36 MeV above threshold, respectively).  

We have carried out calculations for (d,p) reactions populating these four final states in \oonex.  We adopted the recommended spin-parity assignments from Ref.~\cite{Fernandez:11}, namely $5/2^+$ and $1/2^+$ for the ground and first excited states, respectively, and $3/2^+$ for the resonance at  $E_{ex}$ = 4.77 MeV. Since there is some ambiguity as to whether the resonance at $E_{ex}$ = 6.17 MeV has $3/2^+$ or $7/2^-$ character, we considered both possibilities. 
As in the Zr and Ca cases, we carried out full calculations and reproduced the measured cross sections.  We then studied the contributions from the interior post, surface, and exterior prior terms as a function of the separation radius $a$.  

For (d,p) transfers to the two bound states, we found results (not shown here) that are qualitatively similar to those shown in Figure~\ref{fig_Zr_bsPostPriorSurface_3States} for the Zirconium case. The surface terms peak around $a$ = 5 fm and the shape of the full cross section is reproduced by the surface term only, with the overall magnitude being too small by about 20-40\%.

\begin{figure}[htb]
\begin{center}
\includegraphics[width=0.7\columnwidth,angle=270,trim=80 25 20 70,clip=true]{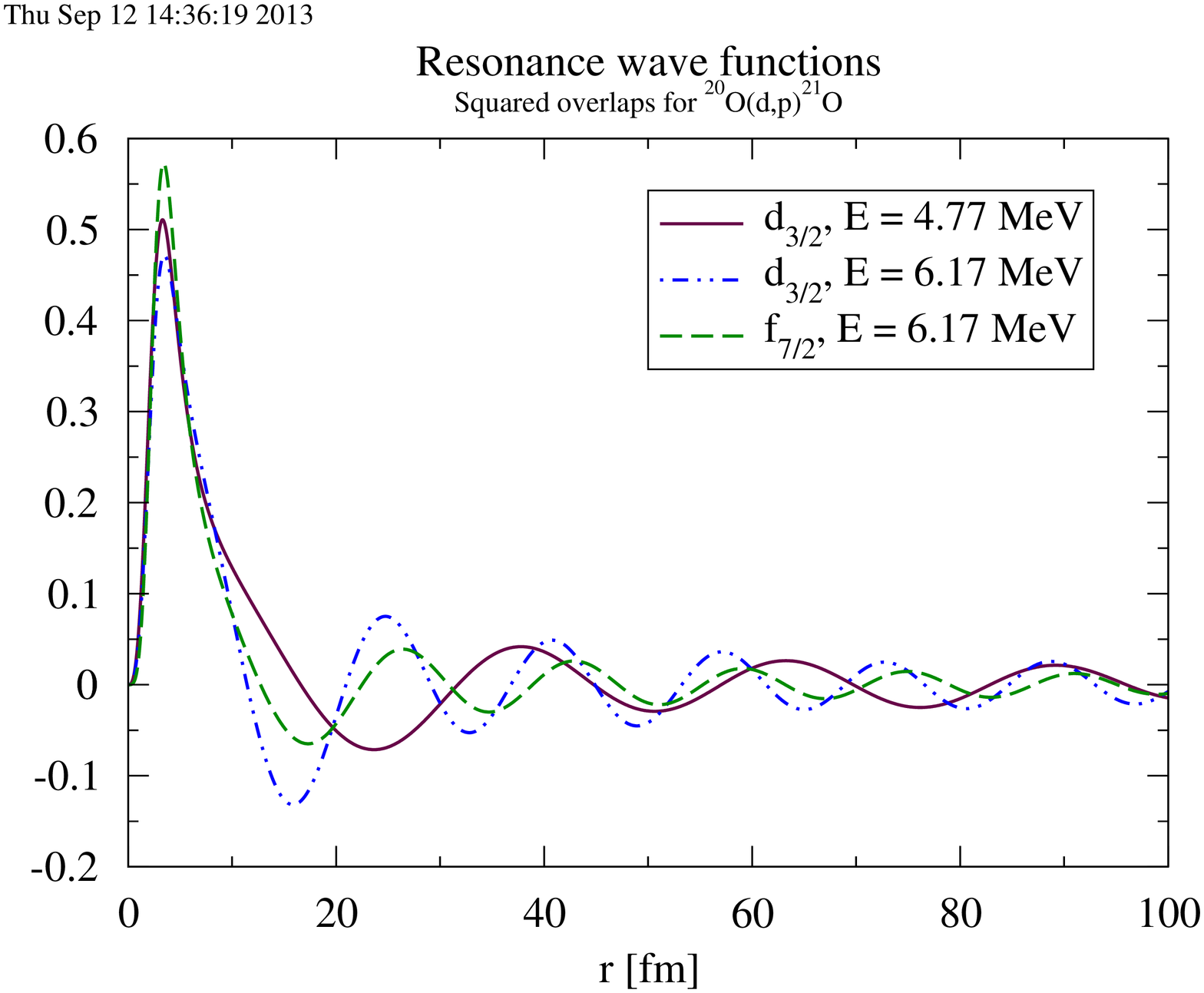}
\end{center}
\caption{
(Color online)
Radial behavior of the one-neutron overlap functions associated with three resonances in \oone (in fm$^{-1}$).  The resonance energies are indicated in the figure.  For the resonance at 6.17 MeV we consider two possible spin-parity assignments.
}
\label{fig_O_resWFs}
\end{figure}

\begin{figure}[htb]
\begin{center}
\includegraphics[width=0.7\columnwidth,angle=270,trim=80 25 30 70,clip=true]{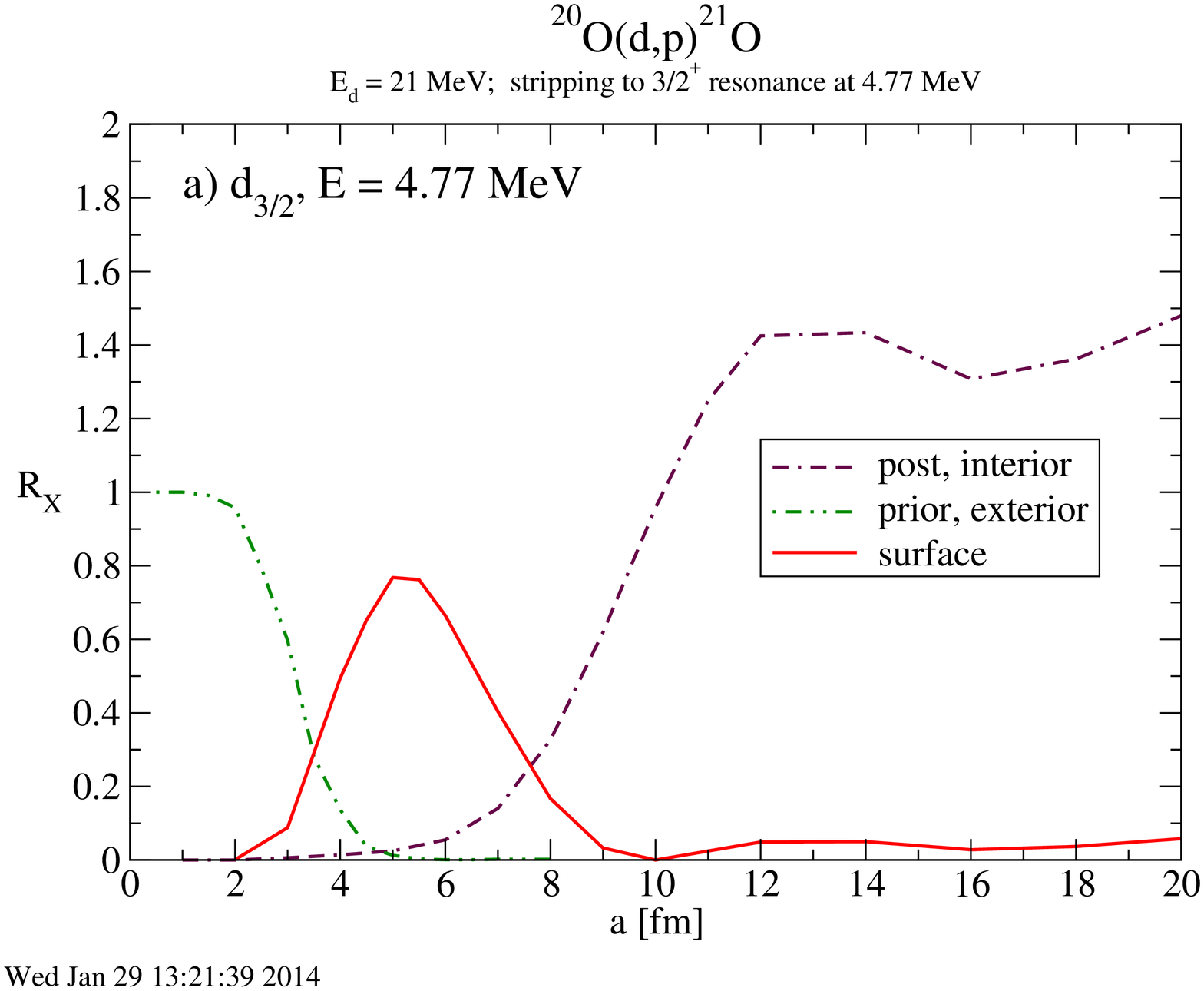}
\includegraphics[width=0.7\columnwidth,angle=270,trim=80 25 30 70,clip=true]{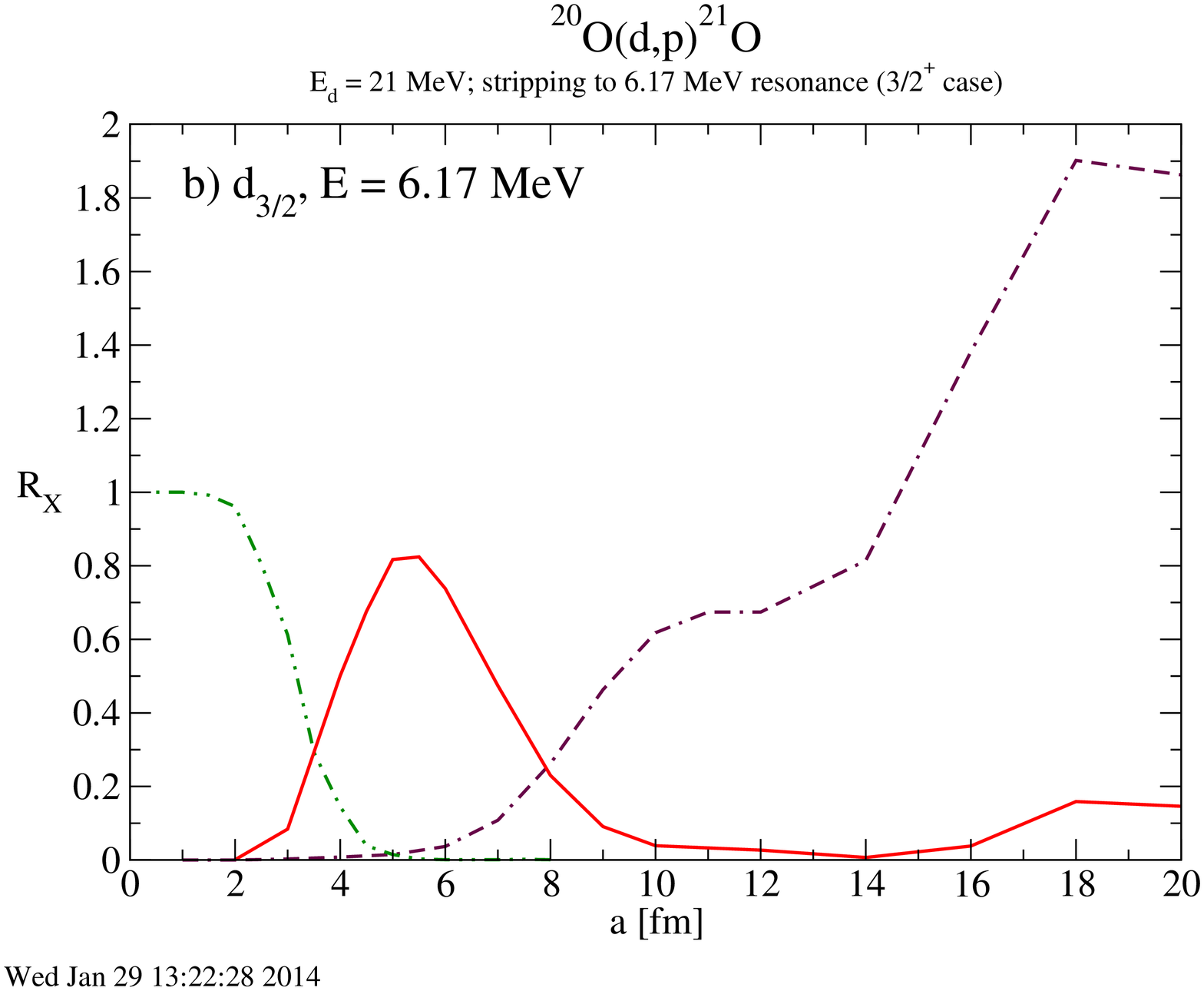} 
\includegraphics[width=0.7\columnwidth,angle=270,trim=80 25 30 70,clip=true]{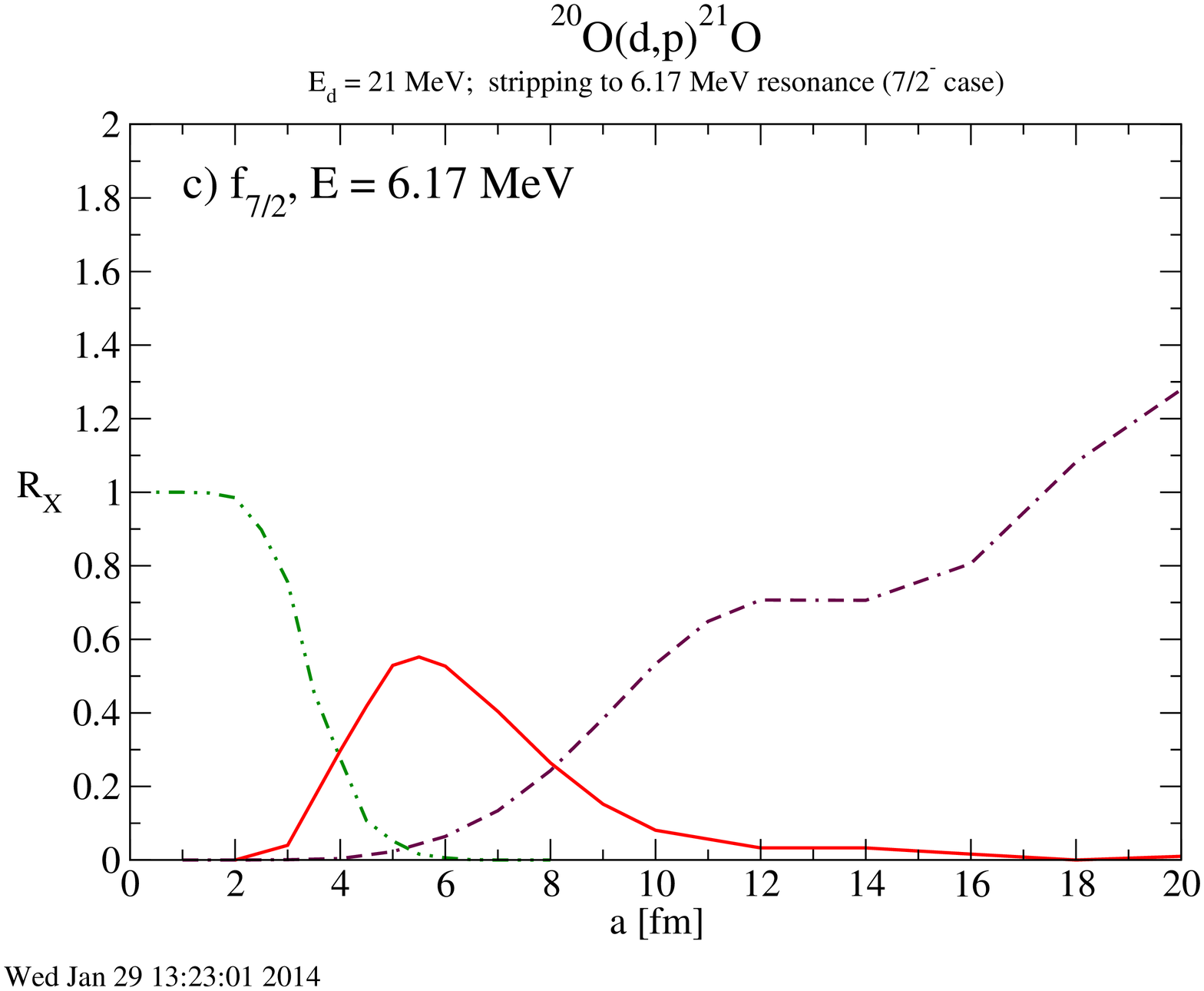} 
\end{center}
\caption{
(Color online)
Examination of interior, surface, and exterior contributions for transfers to resonance states in \oonex. Contributions resulting from the interior (post) term, the surface term, and the exterior (prior) term, are given as a function of the surface radius.
Cross section results are shown for the $3/2^+$ resonance at 4.77 MeV (a) and the 6.17 MeV resonance, assuming a $3/2^+$ state (b), or a $7/2^-$ (c). The associated overlap functions are shown in Figure~\ref{fig_O_resWFs}.
The potential $V_{nA}$, which binds the neutron to the  \otwen nucleus, has a radius of 3.39 fm and a diffuseness of 0.65 fm.
}
\label{fig_O_surfIntExt_3Res}
\end{figure}

Calculations for the resonances were more challenging. This was expected, since these were carried out in the traditional DWBA approach that does not make use of any simplifications that the surface approach might offer.  The one-neutron overlap functions associated with the resonances exhibit oscillations that decline in amplitude only very slowly with increasing radial coordinate, as seen in Figure~\ref{fig_O_resWFs}. As a result, there are contributions to the post-form matrix element (Eq.~\ref{eq:MPost}) from far outside the nucleus that have to be taken into account, while contributions from large distances to the prior-form matrix element (Eq.~\ref{eq:MPrior}) are suppressed by the transition operator. Naturally, this does not only cause problems for the full calculations (when carried out in the post formalism), but also for the calculations of the post-interior contributions, as can be seen in Figure~\ref{fig_O_surfIntExt_3Res}. Even for separation radii $a = $ 20 fm, we find that the ratio $R_{M_{int}(0,a)}$ for the interior-post calculation does not approach unity. To reach this limit, integration out to very large radii (on the order of 100 fm) is required.
The prior calculation does not require integration to such large distances, as the interactions in $\Delta V_{dA}$ (Eq.~\ref{eq_transOp_post}) cease to contribute to the transition matrix element at much smaller radii. 
The exact cross sections shown here were calculated in the prior form. 

Despite the convergence challenges, a region around $r_{nA}$ $\approx 4-7$ fm can be identified for which both prior-exterior contributions and post-interior contributions are small. We calculated the surface term and the associated (d,p) cross sections for several values of $a$ in this region, to test to which extent the surface term is able to reproduce the full cross section.  As in the Zr and Ca cases, we find that the cross sections calculated from the surface term only do not reproduce the full cross sections. The discrepancies are an indication that the remaining terms cannot be completely neglected. 

The comparisons of the surface cross sections with the full cross sections are shown in panels b, d, and f of Figure~\ref{fig_O_res_surfXsecPlus_3Res} in the next section.
There we also discuss an option for improving on the surface-term-only approximation to the (d,p) stripping cross section.

\clearpage
\subsection{Improvements to the surface-integral approach}
\label{sec_results_improvements}

In the previous section we studied the contributions from the interior-post, surface, and exterior-prior terms to the (d,p) cross sections for several target nuclei.  In all cases, for both bound and resonance final states, we found that the surface term gives the dominant contributions, provided a separation radius is chosen that is in the region of the nuclear surface. When comparing to exact calculations of the cross sections, however, we also found that significant strength is missing, which indicates that the residual terms cannot be neglected. In the region where the surface cross section peaks, we find contributions from both the interior-post and the exterior-prior terms.  

A possible path forward for practical applications is to select a separation radius $a$ that is slightly smaller than the radius corresponding to the peak of the surface term.  This will minimize contributions from the post-interior term, thus removing the need for a model for the one-nucleon overlap function in the nuclear interior.  With a decrease in the surface radius comes an increase in the contribution from the prior-exterior term.  
Taking this term explicitly into consideration is necessary for achieving a proper description of the cross section in the DWBA approach tested here.

\begin{figure*}[htb]
\begin{minipage}[b]{0.48\textwidth} 
\begin{center}
\includegraphics[width=0.7\columnwidth,angle=270,trim=80 25 30 70,clip=true]{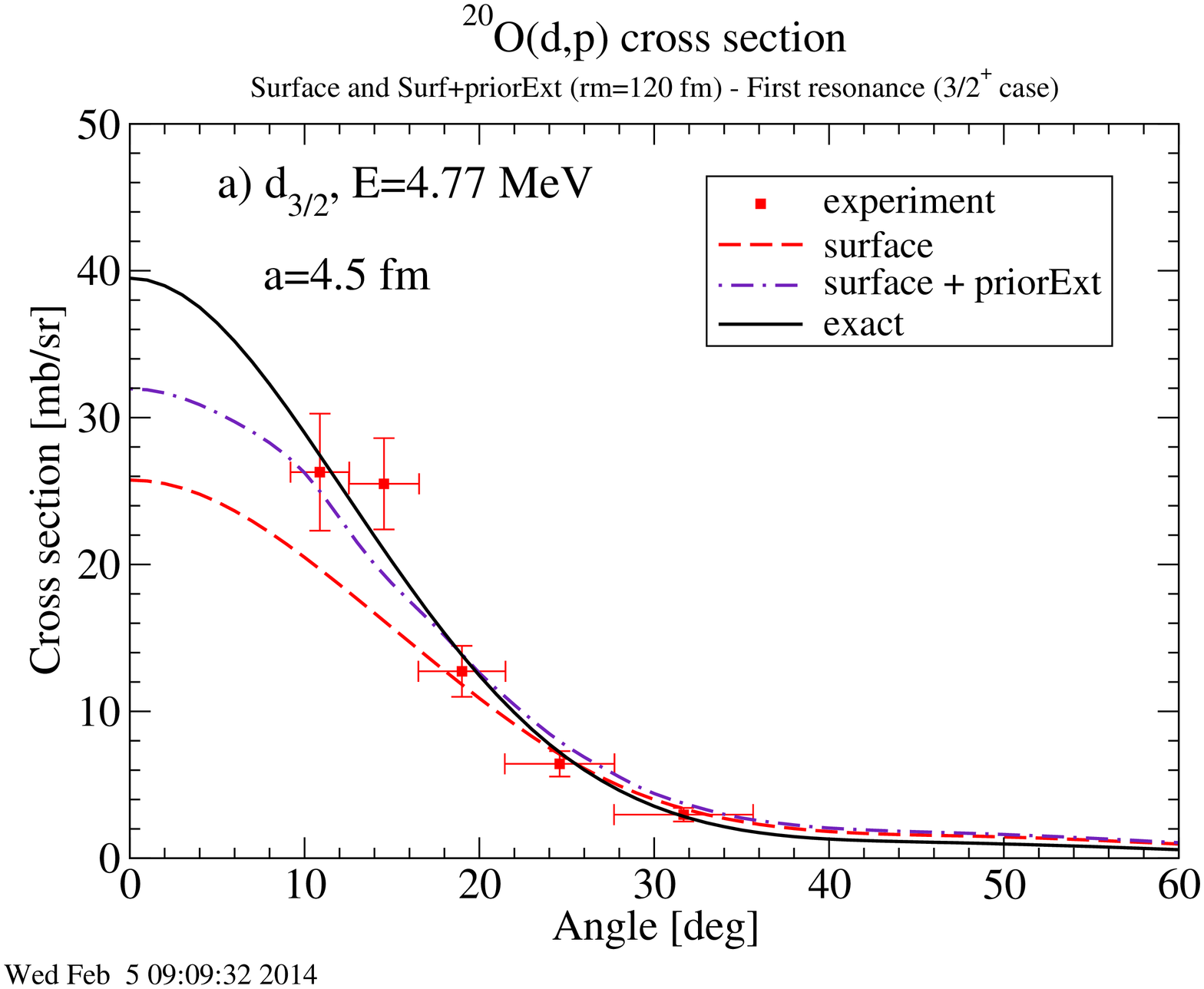}
\includegraphics[width=0.7\columnwidth,angle=270,trim=80 25 30 70,clip=true]{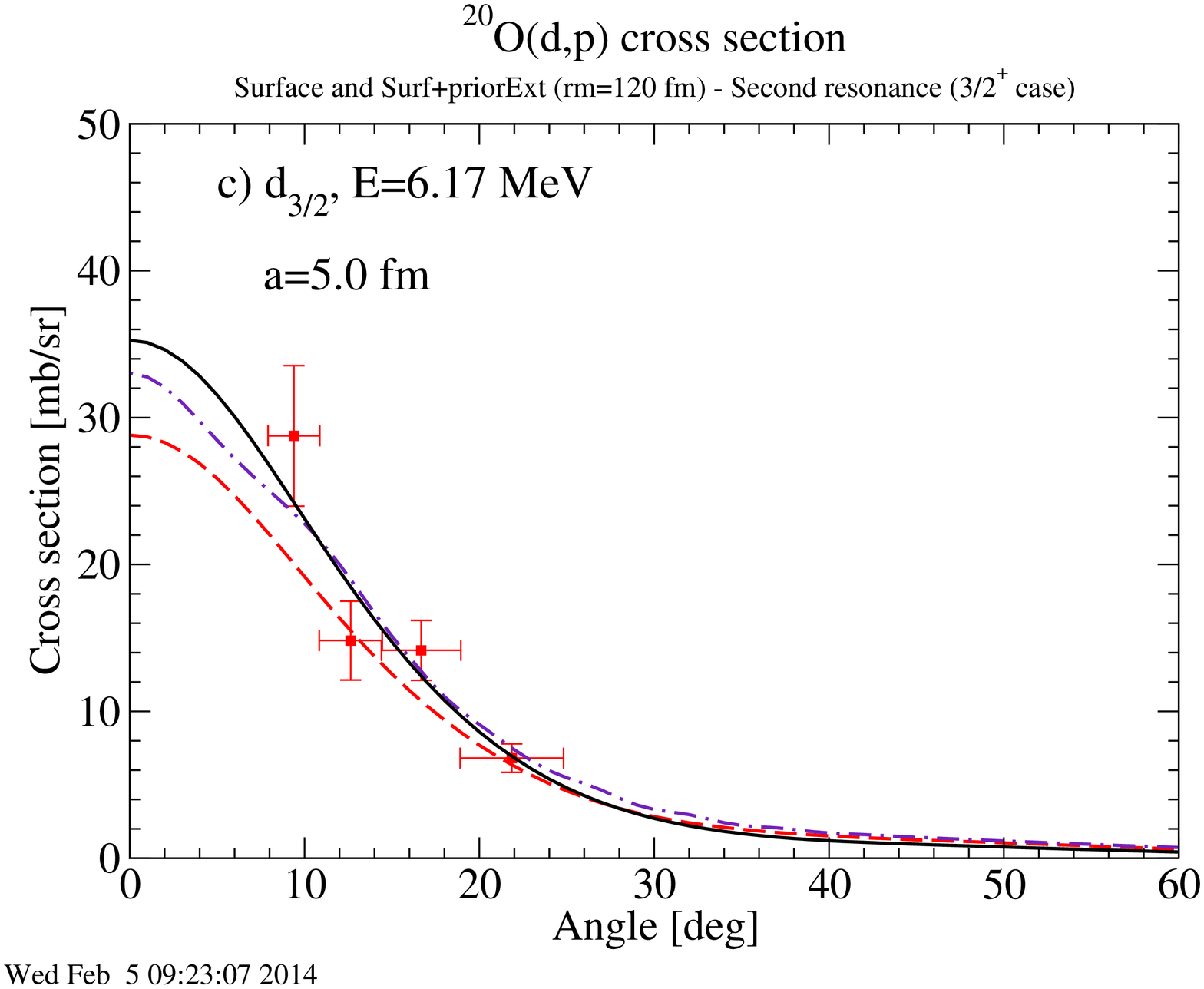} 
\includegraphics[width=0.7\columnwidth,angle=270,trim=80 25 30 70,clip=true]{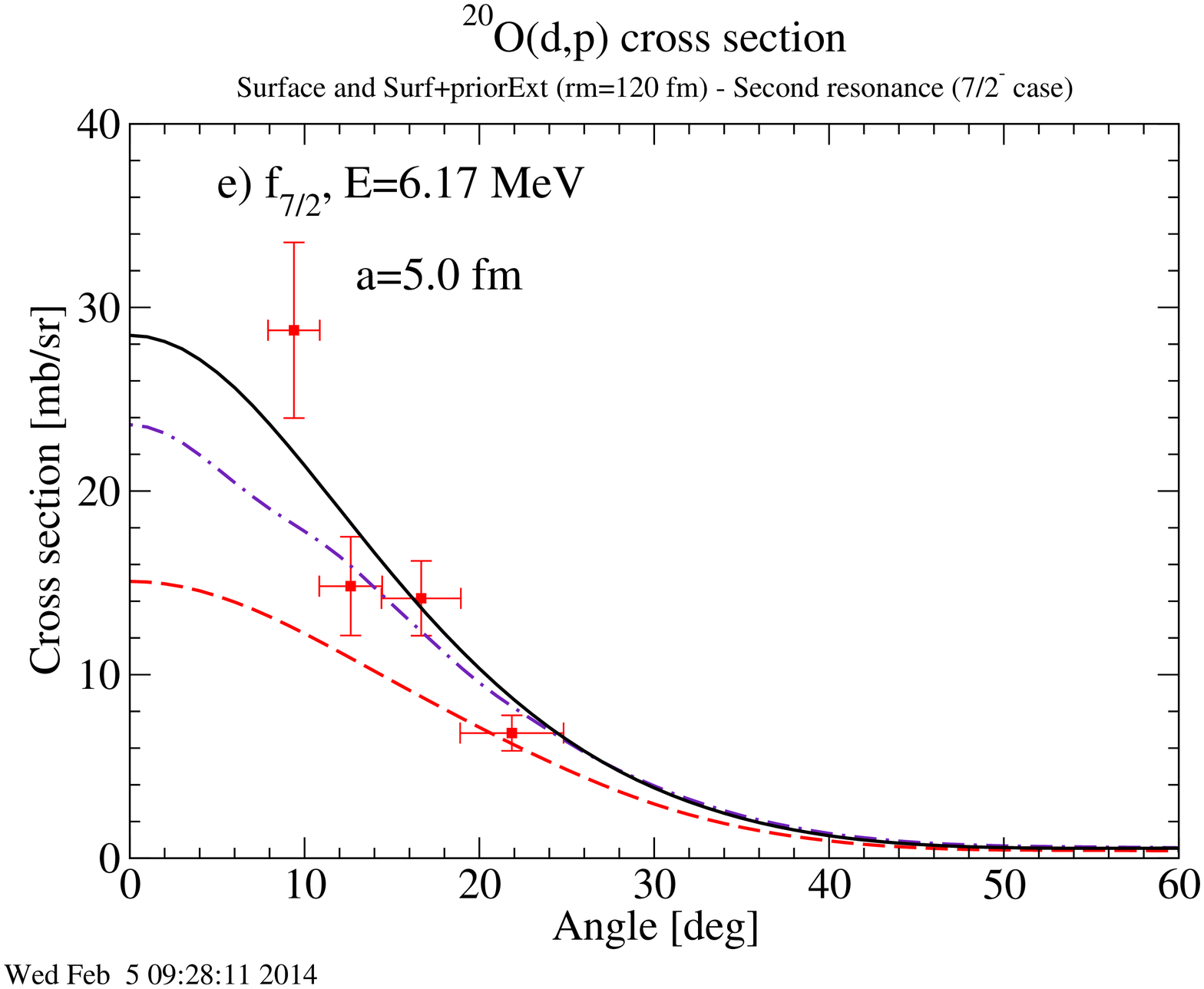} 
\end{center}
\end{minipage}
\hspace{0.1cm}
\begin{minipage}[b]{0.48\textwidth} 
\begin{center}
\includegraphics[width=0.7\columnwidth,angle=270,trim=80 25 30 70,clip=true]{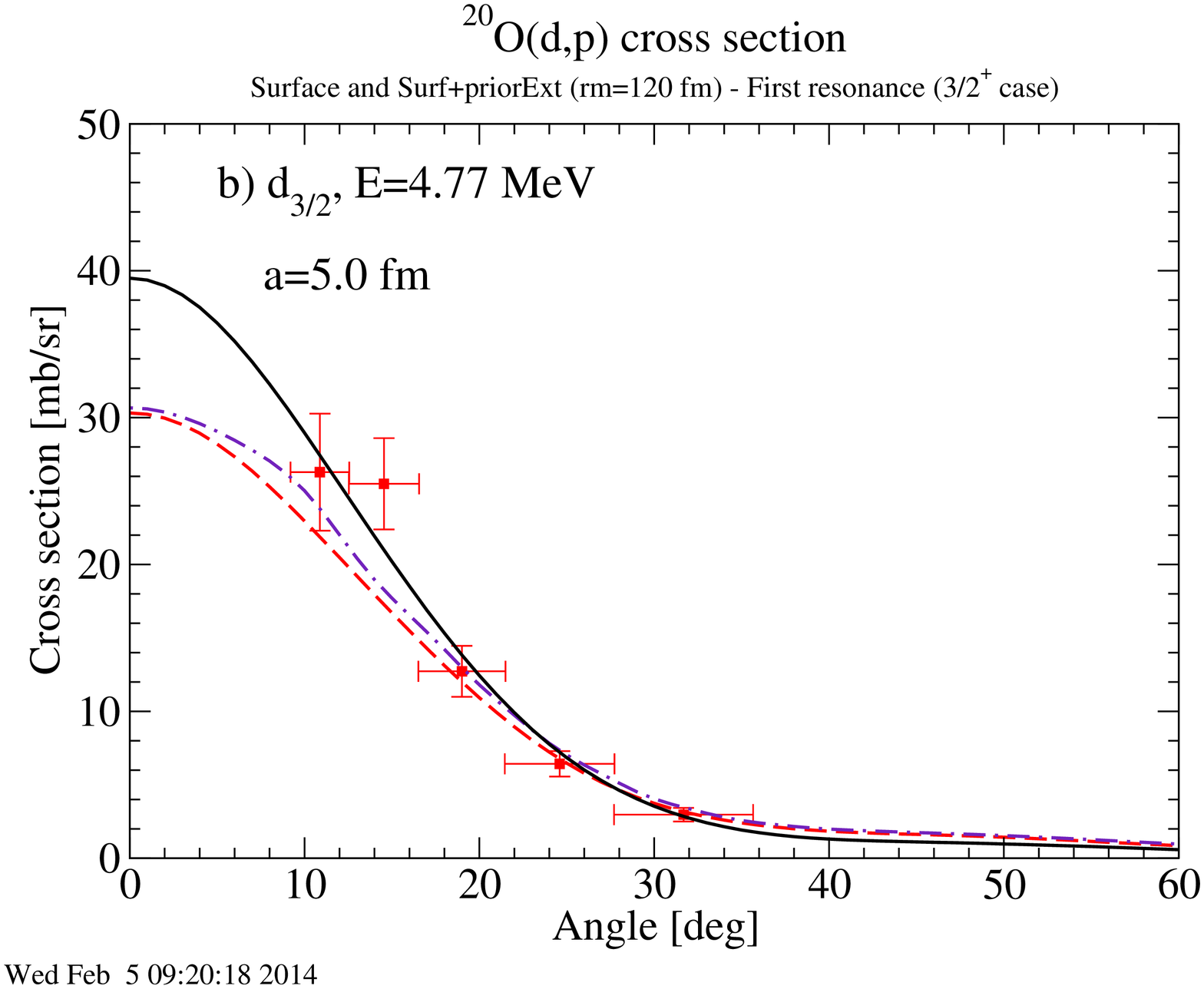}
\includegraphics[width=0.7\columnwidth,angle=270,trim=80 25 30 70,clip=true]{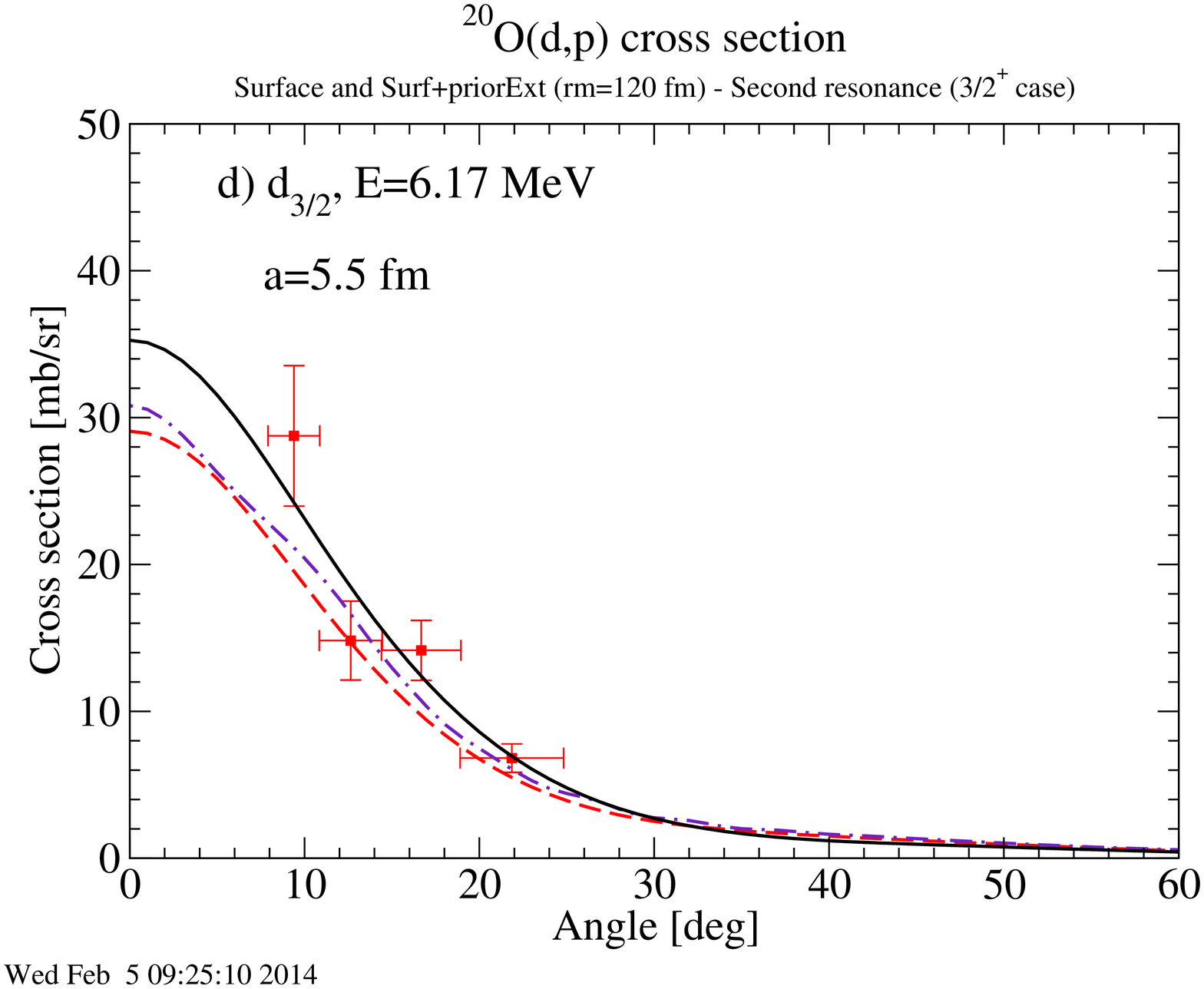} 
\includegraphics[width=0.7\columnwidth,angle=270,trim=80 25 30 70,clip=true]{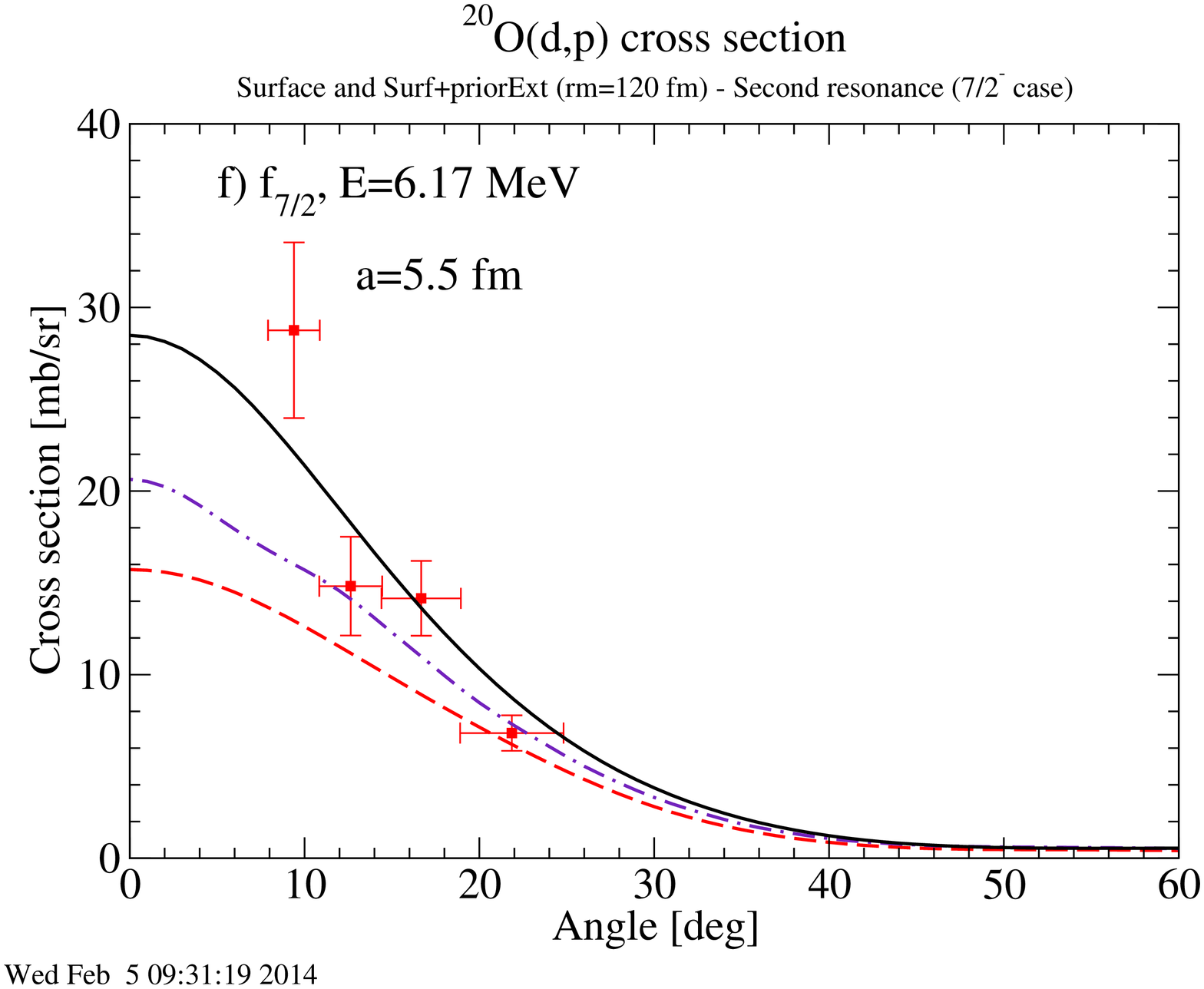} 
\end{center}
\end{minipage}
\caption{
(Color online)
Improvements to the surface-term-only approximation can be achieved by including contributions from the prior-exterior term and selecting a small surface radius.  Panels b), d), f) compare the surface-only results (dashed curve) and the surface plus interior-prior results (dash-dotted curve) to the full calculation (solid line) and to experimental results, for three different resonance cases in \oonex.
The calculations were carried out at surface radii that coincide with the maxima of the surface contributions shown in Figure~\ref{fig_O_surfIntExt_3Res}, specifically b) $a=5.0$ fm for a $3/2^+$ resonance at 4.77 MeV, d) $a=5.5$ fm for a $3/2^+$ resonance at 6.17 MeV, and f) $a=5.5$ fm for a $7/2^-$ resonance at 6.17 MeV.
The left column shows the effect of decreasing the surface radii by 0.5 fm.  Results are shown for the same three resonance cases. Cross sections arising from the surface term are seen to decrease, but cross sections associated with the sum of the surface and the prior-exterior term show better agreement with the exact results. 
}
\label{fig_O_res_surfXsecPlus_3Res}
\end{figure*}

We illustrate the effect of including this term in Figure~\ref{fig_O_res_surfXsecPlus_3Res}, where we compare the surface-only cross sections (dashed curves) and the surface-plus-post-interior calculations (dash-dotted curves) to the exact calculations (solid lines), for the \oone resonances discussed in the previous section.  Available experimental data are shown as well. The surface calculations shown in the right column (panels b, d, f) were calculated with separation radii $a=$ 5.0 fm, 5.5 fm, and 5.5 fm, for a $3/2^+$ resonance at 4.77 MeV, a $3/2^+$ resonance at 6.17 MeV, and a $7/2^-$ resonance at 6.17 MeV, respectively. These choices for the radius $a$ coincide with the maxima of the surface contributions shown in Figure~\ref{fig_O_surfIntExt_3Res}.  As mentioned above, the curves fall clearly short of reproducing the full cross section.
Also shown are calculations that contain both surface and prior-exterior contributions (dash-dotted curves). We observe a slight improvement in the agreement with the exact calculation, but additional contributions (from the post-interior term) would be needed to achieve satisfactory agreement.  

Moving the separation radius to smaller values, however, improves the situation, as a comparison between the right and left columns shows. In panels a), c), e) we compare the surface-only results (dashed curve) and the surface plus interior-prior results (dash-dotted curve) to the full calculation (solid line).  The radii are selected to be 0.5 fm smaller than in the right column. While this shift in $a$ reduces the surface-only cross section, it increases the cross section arising from the the exterior-prior term, with the sum giving a much better approximation to the exact cross section. 
When moving to smaller radii, which improves the agreement of the surface-plus-prior-exterior approximation with the exact result, care must be taken to remain at a radius $a$ for which the nuclear potential $V_{nA}$ can be approximately neglected. In the present case, the potential binding the neutron to the \otwen target has fallen to 15\% of its maximum strength at $a$=4.5 fm and to 8\% of its strength at $a$=5.0 fm. 

The exterior term, which involves an integral over $r_{nA}$ from $a$ to very large distances, includes contributions from the same operator $\Delta V_{dA}$ (see Eq.~\ref{eq:MPrior}) that also causes deuteron breakup. In CDCC calculations, deuteron breakup is explicitly included, through the use of a wave function $\Psi^{CDCC}$ that describes the three-body dynamics between the neutron, proton, and target nucleus.
We thus expect the CDCC implementation of the surface-integral formalism to contain a remnant exterior term that is reduced relative to the DWBA case. This issue remains to be investigated in detail.

Instead of moving to smaller surface radii, one might consider to increase $a$ and include explicitly the contributions from the post-interior term. A test of this strategy (not shown) indicates that this as well leads to improvements in the calculated cross section.
The drawback of this latter approach, however, is that it requires a model for the one-nucleon overlap function in the nuclear interior. It will still be an improvement over the traditional calculations, since i) the interior contribution, and thus the model-dependence, of the cross section is reduced, and ii) the dominant part of the cross section can be parameterized in terms of useful spectroscopic quantities.
We expect these arguments to carry over to the CDCC implementation of the theory. An extension of the present work that will make this possible is underway.

\clearpage

\section{Summary and Outlook}
\label{sec_summary}

Experimentally, resonance structures are most often studied in elastic and inelastic scattering reactions. In this context, the phenomenological R-matrix approach has been extremely useful for the interpretation of experiments and for extracting resonance energies and widths from measured cross sections~\cite{Lane:58,Descouvemont:10}.

A formalism that makes use of R-matrix ideas and is applicable to (d,p) transfer reactions, was recently proposed~\cite{Mukhamedzhanov:11a} with the goal to provide a practical way for extracting structure information from transfer experiments.  
In this approach, the transfer amplitude, a volume integral, was reformulated in terms of a surface integral plus (presumably small) remnant terms that contain contributions from the interior and exterior of the final nucleus. 
The formalism has a series of significant advantages over traditional calculations for cases where the surface-integral dominates the associated cross section.

We investigated the proposed approach for various isotopes in different mass regions, and for a variety of beam energies.  In particular, we studied the separation of the transition matrix element into the three terms suggested in Ref.~\cite{Mukhamedzhanov:11a} and determined their contributions to the total transfer cross sections.

We find that the surface term is dominant in the region where one expects it to be strong, near 5-7 fm for bound states, and at slightly larger radii for resonance states.  At all radii there are also contributions from the post-interior and/or prior-exterior terms.  As a result, the surface term does not completely reproduce the exact cross section. For low to moderate energies (below about $E_d \approx$ 20 MeV), the shape of the cross section is reproduced, but the magnitude differs, by as much as 30-50\%. It is clearly necessary to include additional contributions.  

Possible remedies include increasing the surface radius, which leads to improved cross sections, at the expense of having a model-dependent contribution from the nuclear interior (albeit one that is smaller than the model dependence in traditional calculations).
Alternatively, one can decrease the chosen surface radius.  In the DWBA implementation of the method, this will require the explicit inclusion of external (prior) contributions. In the CDCC extension of the approach, we expect these external contributions to be partially accounted for by the deuteron breakup components that are included in the CDCC formalism.

If the latter strategy is followed, the surface-integral approach has potentially very significant advantages over conventional calculations: 1) It removes the dependence of the cross section calculations on the model used for the nuclear interior. 2) The calculations do not suffer from convergence issues.  3) The method establishes a useful link between resonance properties and transfer observables, since the cross section obtained from the surface integral can be parameterized in terms of quantities that are familiar from traditional R-matrix approaches. 
The strategy is constrained by the requirement that $a$ be outside the nuclear potential $V_{nA}$ and can thus only be employed for peripheral reactions. 

Since the Continuum-Discretized Coupled-Channels (CDCC) formalism describes both transfers and breakup, an extension of the present work is worthwhile exploring.
Work along these lines is underway.



\subsection*{Acknowledgments}
The work is supported through the U.S. Department of Energy's Topical Collaboration TORUS (www.reactiontheory.org).
Is is performed under the auspices of the DOE by Lawrence Livermore National Laboratory under contract DE-AC52-07NA27344, by Ohio University under contracts DE-FG02-93ER40756 and DE-SC0004087, by Michigan State University under contracts DE-FG52-08NA28552 and DE-SC0004084. 
Constructive input from Akram Mukhamdezhanov in the early stages of the project is acknowledged.




\begin{thebibliography}{24}%
\makeatletter
\providecommand \@ifxundefined [1]{%
 \@ifx{#1\undefined}
}%
\providecommand \@ifnum [1]{%
 \ifnum #1\expandafter \@firstoftwo
 \else \expandafter \@secondoftwo
 \fi
}%
\providecommand \@ifx [1]{%
 \ifx #1\expandafter \@firstoftwo
 \else \expandafter \@secondoftwo
 \fi
}%
\providecommand \natexlab [1]{#1}%
\providecommand \enquote  [1]{``#1''}%
\providecommand \bibnamefont  [1]{#1}%
\providecommand \bibfnamefont [1]{#1}%
\providecommand \citenamefont [1]{#1}%
\providecommand \href@noop [0]{\@secondoftwo}%
\providecommand \href [0]{\begingroup \@sanitize@url \@href}%
\providecommand \@href[1]{\@@startlink{#1}\@@href}%
\providecommand \@@href[1]{\endgroup#1\@@endlink}%
\providecommand \@sanitize@url [0]{\catcode `\\12\catcode `\$12\catcode
  `\&12\catcode `\#12\catcode `\^12\catcode `\_12\catcode `\%12\relax}%
\providecommand \@@startlink[1]{}%
\providecommand \@@endlink[0]{}%
\providecommand \url  [0]{\begingroup\@sanitize@url \@url }%
\providecommand \@url [1]{\endgroup\@href {#1}{\urlprefix }}%
\providecommand \urlprefix  [0]{URL }%
\providecommand \Eprint [0]{\href }%
\providecommand \doibase [0]{http://dx.doi.org/}%
\providecommand \selectlanguage [0]{\@gobble}%
\providecommand \bibinfo  [0]{\@secondoftwo}%
\providecommand \bibfield  [0]{\@secondoftwo}%
\providecommand \translation [1]{[#1]}%
\providecommand \BibitemOpen [0]{}%
\providecommand \bibitemStop [0]{}%
\providecommand \bibitemNoStop [0]{.\EOS\space}%
\providecommand \EOS [0]{\spacefactor3000\relax}%
\providecommand \BibitemShut  [1]{\csname bibitem#1\endcsname}%
\let\auto@bib@innerbib\@empty
\bibitem [{\citenamefont {Jones}\ \emph {et~al.}(2010)\citenamefont {Jones},
  \citenamefont {Adekola}, \citenamefont {Bardayan}, \citenamefont {Blackmon},
  \citenamefont {Chae}, \citenamefont {Chipps}, \citenamefont {Cizewski},
  \citenamefont {Erikson}, \citenamefont {Harlin}, \citenamefont {Hatarik},
  \citenamefont {Kapler}, \citenamefont {Kozub}, \citenamefont {Liang},
  \citenamefont {Livesay}, \citenamefont {Ma}, \citenamefont {Moazen},
  \citenamefont {Nesaraja}, \citenamefont {Nunes}, \citenamefont {Pain},
  \citenamefont {Patterson}, \citenamefont {Shapira}, \citenamefont {Shriner},
  \citenamefont {Smith}, \citenamefont {Swan},\ and\ \citenamefont
  {Thomas}}]{Jones:10}%
  \BibitemOpen
  \bibfield  {author} {\bibinfo {author} {\bibfnamefont {K.~L.}\ \bibnamefont
  {Jones}}, \bibinfo {author} {\bibfnamefont {A.~S.}\ \bibnamefont {Adekola}},
  \bibinfo {author} {\bibfnamefont {D.~W.}\ \bibnamefont {Bardayan}}, \bibinfo
  {author} {\bibfnamefont {J.~C.}\ \bibnamefont {Blackmon}}, \bibinfo {author}
  {\bibfnamefont {K.~Y.}\ \bibnamefont {Chae}}, \bibinfo {author}
  {\bibfnamefont {K.~A.}\ \bibnamefont {Chipps}}, \bibinfo {author}
  {\bibfnamefont {J.~A.}\ \bibnamefont {Cizewski}}, \bibinfo {author}
  {\bibfnamefont {L.}~\bibnamefont {Erikson}}, \bibinfo {author} {\bibfnamefont
  {C.}~\bibnamefont {Harlin}}, \bibinfo {author} {\bibfnamefont
  {R.}~\bibnamefont {Hatarik}}, \bibinfo {author} {\bibfnamefont
  {R.}~\bibnamefont {Kapler}}, \bibinfo {author} {\bibfnamefont {R.~L.}\
  \bibnamefont {Kozub}}, \bibinfo {author} {\bibfnamefont {J.~F.}\ \bibnamefont
  {Liang}}, \bibinfo {author} {\bibfnamefont {R.}~\bibnamefont {Livesay}},
  \bibinfo {author} {\bibfnamefont {Z.}~\bibnamefont {Ma}}, \bibinfo {author}
  {\bibfnamefont {B.~H.}\ \bibnamefont {Moazen}}, \bibinfo {author}
  {\bibfnamefont {C.~D.}\ \bibnamefont {Nesaraja}}, \bibinfo {author}
  {\bibfnamefont {F.~M.}\ \bibnamefont {Nunes}}, \bibinfo {author}
  {\bibfnamefont {S.~D.}\ \bibnamefont {Pain}}, \bibinfo {author}
  {\bibfnamefont {N.~P.}\ \bibnamefont {Patterson}}, \bibinfo {author}
  {\bibfnamefont {D.}~\bibnamefont {Shapira}}, \bibinfo {author} {\bibfnamefont
  {J.~F.}\ \bibnamefont {Shriner}}, \bibinfo {author} {\bibfnamefont {M.~S.}\
  \bibnamefont {Smith}}, \bibinfo {author} {\bibfnamefont {T.~P.}\ \bibnamefont
  {Swan}}, \ and\ \bibinfo {author} {\bibfnamefont {J.~S.}\ \bibnamefont
  {Thomas}},\ }\href {http://dx.doi.org/10.1038/nature09048} {\bibfield
  {journal} {\bibinfo  {journal} {Nature}\ }\textbf {\bibinfo {volume} {465}},\
  \bibinfo {pages} {454} (\bibinfo {year} {2010})}\BibitemShut {NoStop}%
\bibitem [{\citenamefont {Volya}\ and\ \citenamefont
  {Zelevinsky}(2006)}]{Volya:06a}%
  \BibitemOpen
  \bibfield  {author} {\bibinfo {author} {\bibfnamefont {A.}~\bibnamefont
  {Volya}}\ and\ \bibinfo {author} {\bibfnamefont {V.}~\bibnamefont
  {Zelevinsky}},\ }\href {\doibase 10.1103/PhysRevC.74.064314} {\bibfield
  {journal} {\bibinfo  {journal} {Phys. Rev. C}\ }\textbf {\bibinfo {volume}
  {74}},\ \bibinfo {pages} {064314} (\bibinfo {year} {2006})}\BibitemShut
  {NoStop}%
\bibitem [{\citenamefont {Mukhamedzhanov}(2011)}]{Mukhamedzhanov:11a}%
  \BibitemOpen
  \bibfield  {author} {\bibinfo {author} {\bibfnamefont {A.~M.}\ \bibnamefont
  {Mukhamedzhanov}},\ }\href {\doibase 10.1103/PhysRevC.84.044616} {\bibfield
  {journal} {\bibinfo  {journal} {Phys. Rev. C}\ }\textbf {\bibinfo {volume}
  {84}},\ \bibinfo {pages} {044616} (\bibinfo {year} {2011})}\BibitemShut
  {NoStop}%
\bibitem [{\citenamefont {Lane}\ and\ \citenamefont {Thomas}(1958)}]{Lane:58}%
  \BibitemOpen
  \bibfield  {author} {\bibinfo {author} {\bibfnamefont {A.~M.}\ \bibnamefont
  {Lane}}\ and\ \bibinfo {author} {\bibfnamefont {R.~G.}\ \bibnamefont
  {Thomas}},\ }\href@noop {} {\bibfield  {journal} {\bibinfo  {journal} {Rev.
  Mod. Phys.}\ }\textbf {\bibinfo {volume} {30}},\ \bibinfo {pages} {257}
  (\bibinfo {year} {1958})}\BibitemShut {NoStop}%
\bibitem [{\citenamefont {Descouvemont}\ and\ \citenamefont
  {Baye}(2010)}]{Descouvemont:10}%
  \BibitemOpen
  \bibfield  {author} {\bibinfo {author} {\bibfnamefont {P.}~\bibnamefont
  {Descouvemont}}\ and\ \bibinfo {author} {\bibfnamefont {D.}~\bibnamefont
  {Baye}},\ }\href@noop {} {\bibfield  {journal} {\bibinfo  {journal} {Reports
  on Progress in Physics}\ }\textbf {\bibinfo {volume} {73}},\ \bibinfo {pages}
  {036301} (\bibinfo {year} {2010})}\BibitemShut {NoStop}%
\bibitem [{\citenamefont {Austern}\ \emph {et~al.}(1987)\citenamefont
  {Austern}, \citenamefont {Iseri}, \citenamefont {Kamimura}, \citenamefont
  {Kawai}, \citenamefont {Rawitscher},\ and\ \citenamefont
  {Yahiro}}]{Austern:87}%
  \BibitemOpen
  \bibfield  {author} {\bibinfo {author} {\bibfnamefont {N.}~\bibnamefont
  {Austern}}, \bibinfo {author} {\bibfnamefont {Y.}~\bibnamefont {Iseri}},
  \bibinfo {author} {\bibfnamefont {M.}~\bibnamefont {Kamimura}}, \bibinfo
  {author} {\bibfnamefont {M.}~\bibnamefont {Kawai}}, \bibinfo {author}
  {\bibfnamefont {G.}~\bibnamefont {Rawitscher}}, \ and\ \bibinfo {author}
  {\bibfnamefont {M.}~\bibnamefont {Yahiro}},\ }\href@noop {} {\bibfield
  {journal} {\bibinfo  {journal} {Phys. Rep.}\ }\textbf {\bibinfo {volume}
  {154}},\ \bibinfo {pages} {125} (\bibinfo {year} {1987})}\BibitemShut
  {NoStop}%
\bibitem [{\citenamefont {Satchler}(1983)}]{Satchler:Book}%
  \BibitemOpen
  \bibfield  {author} {\bibinfo {author} {\bibfnamefont {G.~R.}\ \bibnamefont
  {Satchler}},\ }\href@noop {} {\emph {\bibinfo {title} {Direct nuclear
  reactions}}}\ (\bibinfo  {publisher} {Oxford University Press},\ \bibinfo
  {year} {1983})\BibitemShut {NoStop}%
\bibitem [{\citenamefont {Escher}\ \emph {et~al.}(2001)\citenamefont {Escher},
  \citenamefont {Jennings},\ and\ \citenamefont {Sherif}}]{Escher:01PRC}%
  \BibitemOpen
  \bibfield  {author} {\bibinfo {author} {\bibfnamefont {J.}~\bibnamefont
  {Escher}}, \bibinfo {author} {\bibfnamefont {B.~K.}\ \bibnamefont
  {Jennings}}, \ and\ \bibinfo {author} {\bibfnamefont {H.~S.}\ \bibnamefont
  {Sherif}},\ }\href@noop {} {\bibfield  {journal} {\bibinfo  {journal} {Phys.
  Rev. C}\ }\textbf {\bibinfo {volume} {64}},\ \bibinfo {pages} {065801}
  (\bibinfo {year} {2001})}\BibitemShut {NoStop}%
\bibitem [{\citenamefont {Escher}\ and\ \citenamefont
  {Jennings}(2002)}]{Escher:02PRC}%
  \BibitemOpen
  \bibfield  {author} {\bibinfo {author} {\bibfnamefont {J.}~\bibnamefont
  {Escher}}\ and\ \bibinfo {author} {\bibfnamefont {B.~K.}\ \bibnamefont
  {Jennings}},\ }\href@noop {} {\bibfield  {journal} {\bibinfo  {journal}
  {Phys. Rev. C}\ }\textbf {\bibinfo {volume} {66}},\ \bibinfo {pages} {034313}
  (\bibinfo {year} {2002})}\BibitemShut {NoStop}%
\bibitem [{\citenamefont {Johnson}\ and\ \citenamefont
  {Soper}(1970)}]{Johnson:70}%
  \BibitemOpen
  \bibfield  {author} {\bibinfo {author} {\bibfnamefont {R.~C.}\ \bibnamefont
  {Johnson}}\ and\ \bibinfo {author} {\bibfnamefont {P.~J.~R.}\ \bibnamefont
  {Soper}},\ }\href {\doibase 10.1103/PhysRevC.1.976} {\bibfield  {journal}
  {\bibinfo  {journal} {Phys. Rev. C}\ }\textbf {\bibinfo {volume} {1}},\
  \bibinfo {pages} {976} (\bibinfo {year} {1970})}\BibitemShut {NoStop}%
\bibitem [{\citenamefont {Johnson}\ and\ \citenamefont
  {Tandy}(1974)}]{Johnson:74}%
  \BibitemOpen
  \bibfield  {author} {\bibinfo {author} {\bibfnamefont {R.}~\bibnamefont
  {Johnson}}\ and\ \bibinfo {author} {\bibfnamefont {P.}~\bibnamefont
  {Tandy}},\ }\href {\doibase http://dx.doi.org/10.1016/0375-9474(74)90178-X}
  {\bibfield  {journal} {\bibinfo  {journal} {Nuclear Physics A}\ }\textbf
  {\bibinfo {volume} {235}},\ \bibinfo {pages} {56 } (\bibinfo {year}
  {1974})}\BibitemShut {NoStop}%
\bibitem [{\citenamefont {Michelman}\ \emph {et~al.}(1969)\citenamefont
  {Michelman}, \citenamefont {Fiarman}, \citenamefont {Ludwig},\ and\
  \citenamefont {Robbins}}]{Michelman:69}%
  \BibitemOpen
  \bibfield  {author} {\bibinfo {author} {\bibfnamefont {L.~S.}\ \bibnamefont
  {Michelman}}, \bibinfo {author} {\bibfnamefont {S.}~\bibnamefont {Fiarman}},
  \bibinfo {author} {\bibfnamefont {E.~J.}\ \bibnamefont {Ludwig}}, \ and\
  \bibinfo {author} {\bibfnamefont {A.~B.}\ \bibnamefont {Robbins}},\ }\href
  {\doibase 10.1103/PhysRev.180.1114} {\bibfield  {journal} {\bibinfo
  {journal} {Phys. Rev.}\ }\textbf {\bibinfo {volume} {180}},\ \bibinfo {pages}
  {1114} (\bibinfo {year} {1969})}\BibitemShut {NoStop}%
\bibitem [{\citenamefont {Metz}\ \emph {et~al.}(1975)\citenamefont {Metz},
  \citenamefont {Callender},\ and\ \citenamefont {Bockelman}}]{Metz:75}%
  \BibitemOpen
  \bibfield  {author} {\bibinfo {author} {\bibfnamefont {W.~D.}\ \bibnamefont
  {Metz}}, \bibinfo {author} {\bibfnamefont {W.~D.}\ \bibnamefont {Callender}},
  \ and\ \bibinfo {author} {\bibfnamefont {C.~K.}\ \bibnamefont {Bockelman}},\
  }\href {\doibase 10.1103/PhysRevC.12.827} {\bibfield  {journal} {\bibinfo
  {journal} {Phys. Rev. C}\ }\textbf {\bibinfo {volume} {12}},\ \bibinfo
  {pages} {827} (\bibinfo {year} {1975})}\BibitemShut {NoStop}%
\bibitem [{\citenamefont {Uozumi}\ \emph {et~al.}(1994)\citenamefont {Uozumi},
  \citenamefont {Iwamoto}, \citenamefont {Widodo}, \citenamefont {Nohtomi},
  \citenamefont {Sakae}, \citenamefont {Matoba}, \citenamefont {Nakano},
  \citenamefont {Maki},\ and\ \citenamefont {Koori}}]{Uozumi:94}%
  \BibitemOpen
  \bibfield  {author} {\bibinfo {author} {\bibfnamefont {Y.}~\bibnamefont
  {Uozumi}}, \bibinfo {author} {\bibfnamefont {O.}~\bibnamefont {Iwamoto}},
  \bibinfo {author} {\bibfnamefont {S.}~\bibnamefont {Widodo}}, \bibinfo
  {author} {\bibfnamefont {A.}~\bibnamefont {Nohtomi}}, \bibinfo {author}
  {\bibfnamefont {T.}~\bibnamefont {Sakae}}, \bibinfo {author} {\bibfnamefont
  {M.}~\bibnamefont {Matoba}}, \bibinfo {author} {\bibfnamefont
  {M.}~\bibnamefont {Nakano}}, \bibinfo {author} {\bibfnamefont
  {T.}~\bibnamefont {Maki}}, \ and\ \bibinfo {author} {\bibfnamefont
  {N.}~\bibnamefont {Koori}},\ }\href {\doibase
  http://dx.doi.org/10.1016/0375-9474(94)90740-4} {\bibfield  {journal}
  {\bibinfo  {journal} {Nuclear Physics A}\ }\textbf {\bibinfo {volume}
  {576}},\ \bibinfo {pages} {123 } (\bibinfo {year} {1994})}\BibitemShut
  {NoStop}%
\bibitem [{\citenamefont {Thompson}(1988)}]{Thompson:88}%
  \BibitemOpen
  \bibfield  {author} {\bibinfo {author} {\bibfnamefont {I.~J.}\ \bibnamefont
  {Thompson}},\ }\href@noop {} {\bibfield  {journal} {\bibinfo  {journal}
  {Computer Physics Reports}\ }\textbf {\bibinfo {volume} {7}},\ \bibinfo
  {pages} {167} (\bibinfo {year} {1988})}\BibitemShut {NoStop}%
\bibitem [{\citenamefont {Jr.}(1968)}]{Reid:68}%
  \BibitemOpen
  \bibfield  {author} {\bibinfo {author} {\bibfnamefont {R.~V.~R.}\
  \bibnamefont {Jr.}},\ }\href {\doibase
  http://dx.doi.org/10.1016/0003-4916(68)90126-7} {\bibfield  {journal}
  {\bibinfo  {journal} {Annals of Physics}\ }\textbf {\bibinfo {volume} {50}},\
  \bibinfo {pages} {411 } (\bibinfo {year} {1968})}\BibitemShut {NoStop}%
\bibitem [{\citenamefont {Daehnick}\ \emph {et~al.}(1980)\citenamefont
  {Daehnick}, \citenamefont {Childs},\ and\ \citenamefont
  {Vrcelj}}]{Daehnick:80}%
  \BibitemOpen
  \bibfield  {author} {\bibinfo {author} {\bibfnamefont {W.~W.}\ \bibnamefont
  {Daehnick}}, \bibinfo {author} {\bibfnamefont {J.~D.}\ \bibnamefont
  {Childs}}, \ and\ \bibinfo {author} {\bibfnamefont {Z.}~\bibnamefont
  {Vrcelj}},\ }\href {\doibase 10.1103/PhysRevC.21.2253} {\bibfield  {journal}
  {\bibinfo  {journal} {Phys. Rev. C}\ }\textbf {\bibinfo {volume} {21}},\
  \bibinfo {pages} {2253} (\bibinfo {year} {1980})}\BibitemShut {NoStop}%
\bibitem [{\citenamefont {Koning}\ and\ \citenamefont
  {Delaroche}(2003)}]{Koning:03}%
  \BibitemOpen
  \bibfield  {author} {\bibinfo {author} {\bibfnamefont {A.~J.}\ \bibnamefont
  {Koning}}\ and\ \bibinfo {author} {\bibfnamefont {J.-P.}\ \bibnamefont
  {Delaroche}},\ }\href@noop {} {\bibfield  {journal} {\bibinfo  {journal}
  {Nucl. Phys.}\ }\textbf {\bibinfo {volume} {A713}},\ \bibinfo {pages} {231}
  (\bibinfo {year} {2003})}\BibitemShut {NoStop}%
\bibitem [{\citenamefont {Huby}\ and\ \citenamefont {Mines}(1965)}]{Huby:65}%
  \BibitemOpen
  \bibfield  {author} {\bibinfo {author} {\bibfnamefont {R.}~\bibnamefont
  {Huby}}\ and\ \bibinfo {author} {\bibfnamefont {J.~R.}\ \bibnamefont
  {Mines}},\ }\href {\doibase 10.1103/RevModPhys.37.406} {\bibfield  {journal}
  {\bibinfo  {journal} {Rev. Mod. Phys.}\ }\textbf {\bibinfo {volume} {37}},\
  \bibinfo {pages} {406} (\bibinfo {year} {1965})}\BibitemShut {NoStop}%
\bibitem [{\citenamefont {Vincent}\ and\ \citenamefont
  {Fortune}(1970)}]{Vincent:70}%
  \BibitemOpen
  \bibfield  {author} {\bibinfo {author} {\bibfnamefont {C.~M.}\ \bibnamefont
  {Vincent}}\ and\ \bibinfo {author} {\bibfnamefont {H.~T.}\ \bibnamefont
  {Fortune}},\ }\href {\doibase 10.1103/PhysRevC.2.782} {\bibfield  {journal}
  {\bibinfo  {journal} {Phys. Rev. C}\ }\textbf {\bibinfo {volume} {2}},\
  \bibinfo {pages} {782} (\bibinfo {year} {1970})}\BibitemShut {NoStop}%
\bibitem [{Note1()}]{Note1}%
  \BibitemOpen
  \bibinfo {note} {For transfer to the $^{91}$Zr ground state, there is a clear
  discrepancy in shape between the full calculations and the cross sections
  calculated with the surface term, with the latter not exhibiting the upturn
  at small angles. In fact, in the region below 10 degrees, the measured cross
  section does not exhibit this upturn. The full calculation is not able to
  reproduce the experimental data correctly without introducing a (somewhat
  artifical) lower cutoff in the integration. This issue has already been
  discussed in the literature, see, e.g., Ref.~\cite {Hodgeson:book}, p.
  456.}\BibitemShut {Stop}%
\bibitem [{\citenamefont {Fortune}\ and\ \citenamefont
  {Vincent}(1969)}]{Fortune:69}%
  \BibitemOpen
  \bibfield  {author} {\bibinfo {author} {\bibfnamefont {H.~T.}\ \bibnamefont
  {Fortune}}\ and\ \bibinfo {author} {\bibfnamefont {C.~M.}\ \bibnamefont
  {Vincent}},\ }\href {\doibase 10.1103/PhysRev.185.1401} {\bibfield  {journal}
  {\bibinfo  {journal} {Phys. Rev.}\ }\textbf {\bibinfo {volume} {185}},\
  \bibinfo {pages} {1401} (\bibinfo {year} {1969})}\BibitemShut {NoStop}%
\bibitem [{\citenamefont {Fern\'{a}ndez-Dom\'{i}nguez}\ \emph
  {et~al.}(2011)\citenamefont {Fern\'{a}ndez-Dom\'{i}nguez}, \citenamefont
  {Thomas}, \citenamefont {Catford}, \citenamefont {Delaunay}, \citenamefont
  {Brown}, \citenamefont {Orr}, \citenamefont {Rejmund}, \citenamefont
  {Labiche}, \citenamefont {Chartier}, \citenamefont {Achouri}, \citenamefont
  {Al~Falou}, \citenamefont {Ashwood}, \citenamefont {Beaumel}, \citenamefont
  {Blumenfeld}, \citenamefont {Brown}, \citenamefont {Chapman}, \citenamefont
  {Curtis}, \citenamefont {Force}, \citenamefont {de~France}, \citenamefont
  {Franchoo}, \citenamefont {Guillot}, \citenamefont {Haigh}, \citenamefont
  {Hammache}, \citenamefont {Lapoux}, \citenamefont {Lemmon}, \citenamefont
  {Mar\'echal}, \citenamefont {Moro}, \citenamefont {Mougeot}, \citenamefont
  {Mouginot}, \citenamefont {Nalpas}, \citenamefont {Navin}, \citenamefont
  {Patterson}, \citenamefont {Pietras}, \citenamefont {Pollacco}, \citenamefont
  {Leprince}, \citenamefont {Ramus}, \citenamefont {Scarpaci}, \citenamefont
  {de~S\'er\'eville}, \citenamefont {Stephan}, \citenamefont {Sorlin},\ and\
  \citenamefont {Wilson}}]{Fernandez:11}%
  \BibitemOpen
  \bibfield  {author} {\bibinfo {author} {\bibfnamefont {B.}~\bibnamefont
  {Fern\'{a}ndez-Dom\'{i}nguez}}, \bibinfo {author} {\bibfnamefont {J.~S.}\
  \bibnamefont {Thomas}}, \bibinfo {author} {\bibfnamefont {W.~N.}\
  \bibnamefont {Catford}}, \bibinfo {author} {\bibfnamefont {F.}~\bibnamefont
  {Delaunay}}, \bibinfo {author} {\bibfnamefont {S.~M.}\ \bibnamefont {Brown}},
  \bibinfo {author} {\bibfnamefont {N.~A.}\ \bibnamefont {Orr}}, \bibinfo
  {author} {\bibfnamefont {M.}~\bibnamefont {Rejmund}}, \bibinfo {author}
  {\bibfnamefont {M.}~\bibnamefont {Labiche}}, \bibinfo {author} {\bibfnamefont
  {M.}~\bibnamefont {Chartier}}, \bibinfo {author} {\bibfnamefont {N.~L.}\
  \bibnamefont {Achouri}}, \bibinfo {author} {\bibfnamefont {H.}~\bibnamefont
  {Al~Falou}}, \bibinfo {author} {\bibfnamefont {N.~I.}\ \bibnamefont
  {Ashwood}}, \bibinfo {author} {\bibfnamefont {D.}~\bibnamefont {Beaumel}},
  \bibinfo {author} {\bibfnamefont {Y.}~\bibnamefont {Blumenfeld}}, \bibinfo
  {author} {\bibfnamefont {B.~A.}\ \bibnamefont {Brown}}, \bibinfo {author}
  {\bibfnamefont {R.}~\bibnamefont {Chapman}}, \bibinfo {author} {\bibfnamefont
  {N.}~\bibnamefont {Curtis}}, \bibinfo {author} {\bibfnamefont
  {C.}~\bibnamefont {Force}}, \bibinfo {author} {\bibfnamefont
  {G.}~\bibnamefont {de~France}}, \bibinfo {author} {\bibfnamefont
  {S.}~\bibnamefont {Franchoo}}, \bibinfo {author} {\bibfnamefont
  {J.}~\bibnamefont {Guillot}}, \bibinfo {author} {\bibfnamefont
  {P.}~\bibnamefont {Haigh}}, \bibinfo {author} {\bibfnamefont
  {F.}~\bibnamefont {Hammache}}, \bibinfo {author} {\bibfnamefont
  {V.}~\bibnamefont {Lapoux}}, \bibinfo {author} {\bibfnamefont {R.~C.}\
  \bibnamefont {Lemmon}}, \bibinfo {author} {\bibfnamefont {F.}~\bibnamefont
  {Mar\'echal}}, \bibinfo {author} {\bibfnamefont {A.~M.}\ \bibnamefont
  {Moro}}, \bibinfo {author} {\bibfnamefont {X.}~\bibnamefont {Mougeot}},
  \bibinfo {author} {\bibfnamefont {B.}~\bibnamefont {Mouginot}}, \bibinfo
  {author} {\bibfnamefont {L.}~\bibnamefont {Nalpas}}, \bibinfo {author}
  {\bibfnamefont {A.}~\bibnamefont {Navin}}, \bibinfo {author} {\bibfnamefont
  {N.}~\bibnamefont {Patterson}}, \bibinfo {author} {\bibfnamefont
  {B.}~\bibnamefont {Pietras}}, \bibinfo {author} {\bibfnamefont {E.~C.}\
  \bibnamefont {Pollacco}}, \bibinfo {author} {\bibfnamefont {A.}~\bibnamefont
  {Leprince}}, \bibinfo {author} {\bibfnamefont {A.}~\bibnamefont {Ramus}},
  \bibinfo {author} {\bibfnamefont {J.~A.}\ \bibnamefont {Scarpaci}}, \bibinfo
  {author} {\bibfnamefont {N.}~\bibnamefont {de~S\'er\'eville}}, \bibinfo
  {author} {\bibfnamefont {I.}~\bibnamefont {Stephan}}, \bibinfo {author}
  {\bibfnamefont {O.}~\bibnamefont {Sorlin}}, \ and\ \bibinfo {author}
  {\bibfnamefont {G.~L.}\ \bibnamefont {Wilson}},\ }\href {\doibase
  10.1103/PhysRevC.84.011301} {\bibfield  {journal} {\bibinfo  {journal} {Phys.
  Rev. C}\ }\textbf {\bibinfo {volume} {84}},\ \bibinfo {pages} {011301}
  (\bibinfo {year} {2011})}\BibitemShut {NoStop}%
\bibitem [{\citenamefont {Hodgeson}(1971)}]{Hodgeson:book}%
  \BibitemOpen
  \bibfield  {author} {\bibinfo {author} {\bibfnamefont {P.~E.}\ \bibnamefont
  {Hodgeson}},\ }\href@noop {} {\emph {\bibinfo {title} {Nuclear Reactions and
  Nuclear Structure}}}\ (\bibinfo  {publisher} {Oxford University Press},\
  \bibinfo {year} {1971})\BibitemShut {NoStop}%
\end{thebibliography}

%

\end{document}